\documentclass[a4paper, amsfonts, amssymb, amsmath, reprint, showkeys, nofootinbib, twoside, superscriptaddress]{revtex4-1}
\usepackage[english]{babel}
\usepackage[utf8]{inputenc}
\usepackage[colorinlistoftodos, color=green!40, prependcaption]{todonotes}
\usepackage{amsmath}
\usepackage{txfonts}
\usepackage{comment}
\usepackage{ulem}

\usepackage{amsthm}
\usepackage{mathtools}
\usepackage{physics}
\usepackage{xcolor}
\usepackage{graphicx}
\usepackage[left=23mm,right=13mm,top=35mm,columnsep=15pt]{geometry} 
\usepackage{adjustbox}
\usepackage{placeins}
\usepackage[T1]{fontenc}
\usepackage{lipsum}
\usepackage{csquotes}

\usepackage{hyperref} 
\begin{document}
\title{Molecularity: a fast and efficient criterion for probing superconductivity}

\author{Mat\'ias E. di Mauro}
    \affiliation{Laboratoire de Chimie Th\'eorique, Sorbonne Universit\'e \& CNRS, 4 Pl. Jussieu, 75005, Paris, France }
\author{Beno\^it Bra\"ida}
     \affiliation{Laboratoire de Chimie Th\'eorique, Sorbonne Universit\'e \& CNRS, 4 Pl. Jussieu, 75005, Paris, France }
\author{Ion Errea}
    \affiliation{Fisika Aplikatua Saila, Gipuzkoako Ingeniaritza Eskola, University of the Basque Country (UPV/EHU), Europa Plaza 1, 20018 Donostia/San Sebastián, Spain}
    \affiliation{Centro de Física de Materiales (CSIC-UPV/EHU), Manuel de Lardizabal Pasealekua 5, 20018 Donostia/San Sebastián, Spain }
    \affiliation{Donostia International Physics Center (DIPC), Manuel de Lardizabal Pasealekua 4, 20018 Donostia/San Sebastián, Spain }
\author{Trinidad Novoa}
    \email[Correspondence email address: ]{trinidad.novoa_aguirre@sorbonne-universite.fr}
    \affiliation{Laboratoire de Chimie Th\'eorique, Sorbonne Universit\'e \& CNRS, 4 Pl. Jussieu, 75005, Paris, France }
\author{Julia Contreras-Garc\'ia}
    \email[Correspondence email address: ]{contrera@lct.jussieu.fr}
    \affiliation{Laboratoire de Chimie Th\'eorique, Sorbonne Universit\'e \& CNRS, 4 Pl. Jussieu, 75005, Paris, France }

\date{\today} % Leave empty to omit a date

\begin{abstract}
We present an efficient criterion for probing the critical temperature of hydrogen based superconductors. 
We start by expanding the applicability of 3D descriptors of electron localization 
to superconducting states within the framework of superconducting DFT. We first apply this descriptor to a model system, the hydrogen chain, which allows to prove two main concepts: i) that the electron localization changes very little when the transition from the normal to the superconducting state takes place, i.e. that it can be described at the DFT level from the normal state; and ii) that the formation of molecules can be characterized within this theoretical framework, enabling to filter out systems with marked molecular character and hence with low potential to be good superconductors. These two ideas, are then exploited in real binary and ternary systems, showing i) that the bonding type can be characterized automatically; and ii) that this provides a new index which enables to feed machine learning algorithms for a better prediction of critical temperatures. Overall, this sets a grounded theoretical scenario for an automatic and efficient high-throughput of potential hydrogen based superconductors.

\end{abstract}

\keywords{superconductivity, electron localization, molecularity}

\maketitle

%\section{Introduction}

Superconductivity can be considered among the most exciting discoveries in material science of the XXth century due to its implications both at the technological and scientific levels. These implications have led to the discovery of a plethora of superconducting families
%an explosion of superconductor searches in the last decades 
to which the high pressure hydrides have been added in the last years.

Few hydrate examples are H$_3$S \cite{DrozdovEremets_Nature2015}, YH$_9$ \cite{Kong2021}, YH$_6$ \cite{Troyan2021}, and LaH$_{10}$ \cite{Drozdov2019}; all reaching critical temperatures (T$_c$) well above 200 K at megabar pressures. 
%In addition, there have been several studies claiming room-temperature superconductivity, which have not been free of controversy. Among them there is the alleged observation of a T$_c$ of 288 K in a compound formed by sulfur, carbon, and hydrogen \cite{Snider2020}, and of $T_c=294$ K for a Lu-N-H compound \cite{Dasenbrock2023}.
However, the hard conditions for the synthesis as well as the difficult experimental characterization, make the statement of new high $T_c$ materials difficult from the experimental viewpoint. 
In this panorama, theory has become a trustworthy diagnosis of hydride superconductivity. 
Nevertheless, this comes at a high computational cost.
As an example, the calculation of the T$_c$ of LaH$_{10}$ within the anharmonic approximation takes hundreds of thousands of CPU hours. 

Given the high critical temperatures that hydrides have shown to have,
the search of high T$_c$ hydride superconductors is still well alive, but it is claiming for a more efficient theoretical approach.

A faster alternative would be to find cheap footprints of superconductivity from the electronic structure.
If we collect the main characteristics that have been put together over the years, we have some chemical/structural features, e.g. hydrogen-rich systems mixed with s and p elements and highly symmetrical structures favour high T$_c$.
From the electronic structure viewpoint, materials with a high density of states (DOS) at the Fermi level are the best candidates for high-temperature superconductivity \cite{Belli21}. 
Looking at the normal state-superconductor transition, it has been proposed that the mechanism of superconductivity can be traced to the the formation of electronic pairs be it in the shape of strongly covalent metallic bonds (MgB$_2$) \cite{mgb2} or lone pairs (Te)\cite{te}. 

But, even if these features can suggest good trends, they provide necessary but not sufficient conditions. Trying to find a sufficient condition, some of the authors have recently shown that a correlation exists between the critical temperature and the delocalization channels at the density functional theory (DFT) level \cite{Belli21}. These channels are determined thanks to the Electron Localization Function (ELF).
The ELF \cite{Becke1990,SavinELF} (see S.I. for more details) allows to identify regions of localized same-spin electron pairs. Its core, $\chi$, can be understood as a relative measure of the excess of local kinetic energy density (KED) of fermions with respect to bosons:
\begin{equation}
\chi(r)=\frac{\tau(r)-\tau_{vW}(r)}{\tau_{TF}(r)}\, ,
\label{tp}
\end{equation}
where $\tau(r)$,  is the kinetic energy density of the system, and $\tau_{vW}(r)$ \cite{vonWeizsacker1935} is its form in the von-Weizsacker approximation. For a 3D system, the Thomas-Fermi kinetic energy density takes the form $\tau_{TF}(r)=3/10(3\pi^2)^{2/3}\rho(r)^{5/3}$. Note that in the case of a 1D system -as used in part of this contribution- it becomes $\tau_{TF} (x) =\pi^2/24\, \rho(x)^3$ \cite{Trappe2023}.
For representation reasons, $\chi$ is usually inversed and rescaled, leading to the common expression of ELF (also named $\eta$), which varies in the [0,1] range: 
\begin{align}
    \eta(r) = \left[1+ \chi(r)^2\right]^{-1}\, .
\end{align}
Because the von-Weizs\"acker kinetic energy is exact for a bosonic system of the same density $\rho(r)$, the term $\chi$ is a local measure of the excess kinetic energy due to the fermionic nature of the electrons, also known as Pauli kinetic energy density. If this quantity is high, it means that electron pairs are delocalized in that region, and the ELF function value will be small. If the kinetic energy density is not locally increased as an effect of the Pauli exclusion principle, we say that electrons are localized, which will be reflected on a high value of the ELF. 

Some of the authors have shown that ELF can be used to classify bonds in binary supercoductors in six distinct families: molecular systems, covalent systems, systems influenced by weak covalent hydrogen-hydrogen interactions, systems exhibiting electride behavior, ionic systems, and isolated systems. In each instance, the bond nature is identified through analyzing ELF saddle points between different atoms.
Moreover, we also found that the value of the ELF at the saddle point which leads to a surface revealing a 3D delocalization through the cell (hereafter called the "networking value") correlates with the critical temperature of superconductors \cite{Belli21}.

Nevertheless, this initial proposal was not absent of limitations. The fact that DFT calculations can be used for describing the onset of superconductivity needs to be understood. The inherent nature of the networking value, although intuitive, needs to be further explored. This is so much so, if we take into account that for a high throughput exploitation of this index, it is necessary to make sure that it is applicable to more complex systems (ternary, quaternaries, etc) and that no information is missing in the correlation.

These points will be addressed here. Firstly, we will dwell on the use of DFT for describing electron localization in superconductors. With this aim in mind, we will develop a new formulation of the ELF within the superconducting DFT framework and we will apply it to a model system. This will allow us to prove that the normal state DFT analysis of the ELF is sufficient to describe that of the superconducting state. %This will allow us to prove that indeed DFT level is enough for a first approach at the ELF values.
It will also allow us to identify other bonding features, such as the formation of molecules, which quantitatively characterize bad superconductors. With these tools at hand, we will prove in a set of binary and ternary compounds that the new index allows to i) complement the networking value in more complex systems, and ii) improve the fast prediction of $T_c$; with an special focus in high $T_c$ superconductors (more difficult to predict due to the lack of data).
%Given this correlation, electron localization as measured by the ELF function at the DFT level seems to be able to capture the chemical onset of superconductivity. As we will see in the following, this feature of the ELF function enabled us to propose a fast and simple criterion based on a standard electronic structure DFT calculation, to identify materials likely to display high critical temperature superconductivity.

%\textcolor{red}{Why is all these necessary? Maybe the question to formulate is how can a property of the normal state give info about the superconducting nature?  a paragraph exposing the problem and later saying "In this letter we show..."}

%\textcolor{red}{aim summary}

%\textcolor{red}{The motivation to study the hydrogen chain should be better clarified., after the main goal and the hypothesis are outlined before.}

%\section{ELF in the superconducting state}

In the first place, in order to be able to understand the connection between the DFT description of the normal and the superconducting state, we will resort to superconducting DFT (SC-DFT) \cite{gross,gross2,Oliveira1988, Sanna17,sanna}, which has been shown to very accurately reproduce experimental $T_c$'s of conventional
superconductors without introducing any empirical parameters.
SC-DFT is an adapted formulation of DFT for superconductors, where on-top of the common external potential that couples to electrons, the Hamiltonian comprises external potentials taking into account the most important characteristics of superconductors: electron-phonon coupling and symmetry breaking allowing Cooper pairs to tunnel in and out of the system (see S.I. for more information on SC-DFT).

Within this approach the Hamiltonian includes a non-particle conserving superconducting symmetry breaking term and the theory is formulated in the grand-canonical ensemble.  
The electron density in the superconducting state becomes (see S.I. for the demonstration):
\begin{align}
    \rho^{SC}(r) = \sum_{nk} \left[1 -\frac{\xi_{nk}}{\abs{E_{nk}}}\tanh{\left(\frac{\beta\abs{E_{nk}}}{2}\right)} \right]\abs{\varphi_{nk}(r)}^2\, ,
    \label{eq:scdft-rho-final}
\end{align}
where we have used the \emph{SC} superscript to differentiate this density from the normal state electron density.  Note that within this approach the  orbitals energies $E_{nk}$ take the form  $E_{nk} = \sqrt{\xi_{nk}^2 + \Delta_s(nk)^2} $, with $\Delta_s(nk)$ being the superconducting gap, and $\varphi_{nk}(r)$ and $\xi_{nk}$ the Kohn-Sham orbitals and energies of the periodic system, respectively.
Note that in the normal state limit, $\Delta_s(nk) \to 0$, so that we recover the normal state density,
\begin{align}
    \rho^{\rm NS}(r) &= 2 \sum_{nk} f(\xi_{nk}) \abs{\varphi_{nk}(r)}^2\, ,
    \label{eq:scdft-rho-NS}
\end{align}
where $f(E_i)=(1+e\,^{\beta E_i})^{-1}$ is the Fermi-Dirac distribution. This point is crucial for the analysis of the transition.

The SC-DFT framework allows us to define a generalized superconducting ELF. We define the superconducting one-reduced density matrix (1-RDM), $\rho^{SC}_1(r,r')$ (see S.I. for a full development):
\begin{align}
\rho_1^{\rm SC}(r,r') = \sum_{nk} n^{SC}_{nk} \varphi^*_{nk}(r)\varphi_{nk}(r') \, ,
\label{eq:1RDM-final}
\end{align}
where we have writen the SC 1-RDM in terms of Kohn-Sham orbitals and SC occupations:
\begin{align}
n^{SC}_{nk} \equiv 1 -\frac{\xi_{nk}}{\abs{E_{nk}}}\tanh{\left(\frac{\beta\abs{E_{nk}}}{2}\right)} \, .
\end{align}
Notice that in the normal state limit, we recover
\begin{align}
    \rho_1^{\rm NS}(r,r') &= 2 \sum_{nk} f(\xi_{nk}) \varphi^*_{nk}(r)\varphi_{nk}(r')
\end{align}
It can be seen that within the SC-DFT framework, the difference of the 1-RDM of both states is only reflected in a change of the occupation numbers. 

With the density matrix in \eqref{eq:1RDM-final}, we can compute the kinetic energy of the system,
\begin{align}
    T^{SC} &= -\frac{1}{2} \int\left.\nabla_{r'}^2\rho^{SC}_1(r,r')\right\rvert_{r'=r}\, dr\, , \\
%    &= -\frac{1}{2}  \sum_{nk} n^{SC}_{nk}\int \varphi^*_{nk}(r)\nabla^2\varphi_{nk}(r) \, dr\, , \\
    &= \frac{1}{2}  \sum_{nk}  n^{SC}_{nk} \int \abs{\nabla\varphi_{nk}(r)}^2 \, dr\, , 
\end{align}
letting us define a positive definite kinetic energy density,
\begin{align}
\tau^{SC}(r) \equiv \frac{1}{2}  \sum_{nk}  n^{SC}_{nk} \abs{\nabla\varphi_{nk}(r)}^2\, .
\label{eq:KED}
\end{align}

We can also obtain the von Weizsäcker and Thomas-Fermi KEDs,
\begin{align}
\label{eq:KED_vW}
\tau^{SC}_{vW}(r) &= \frac{1}{8} \frac{\abs{\nabla\rho(r)}^2}{\rho(r)}\, ,\\
\tau^{SC}_{TF}(r) &= \frac{3}{10} (3\pi^2)^{2/3}\, \rho(r)^{5/3}\,. 
\label{eq:KED_TF}
\end{align}
Note that here $\tau^{SC}_{TF}(r)$ is defined in 3D. Finally, equations \eqref{eq:KED}, \eqref{eq:KED_vW} and \eqref{eq:KED_TF} allow us define the ELF for the superconducting state as:
\begin{align}
\eta^{SC}(r) = \left[1+ \left( \frac{\tau^{SC}(r)-\tau^{SC}_{vW}(r)}{\tau^{SC}_{TF}(r)}\right)^2\right]^{-1}\, .
\label{eq:SCELF}
\end{align}
As all the superconducting quantities defined in this framework, the superconducting ELF (SC-ELF) converges to the (temperature-dependent) normal state ELF when the gap goes to zero.

Note that in order use to the expression in eq.~\ref{eq:SCELF} to examine the SC-ELF in a model or in a real system, it is necessary to have an expression for the gap. In SC-DFT, this is done making use of the Green's functions formalism. It is possible, however, to introduce an approximation and represent the dependence of the gap at zero Kelvin with respect to the energies $\xi$ as an isotropic Lorentzian function \cite{Tinkham1996},
\begin{align}
    \Delta_0(\xi) = \frac{\Delta_0}{N_0 \pi}~\frac{\omega/2}{\xi^2+(\omega/2)^2} \, ,
\end{align}
where $\omega$ is a parameter that adjusts the width of the peak, and $N_0$ is a normalization such that the height of the peak at $\xi=0$ is $\Delta_0$. The latter is the constant of the gap at $T=0$ K in BCS \cite{Bardeen1957}, that depends on the critical temperature, $\Delta_0 = 1.76 k_B T_c$,
with $k_B$ being Boltzmann's constant. Then, considering the dependence of the gap with respect to the temperature \cite{Tinkham1996, Kajimura91}, 
we shall use
\begin{align}
    \Delta(\xi;T) = \Delta_0(\xi) \tanh{\left(1.74\sqrt{\frac{T_c-T}{T}}\right)}\, .
    \label{eq:Delta-xi-T}
\end{align}
so that in our model, the gap can be obtained for any temperature and energy, for a given $T_c$.

%\section{Delocalization in a 1D model chain}

In order to analyze the behavior of SC-ELF we will first apply it to a simple model, the 1D hydrogen chain. This calls for the one-dimensional definition of $\tau_{TF}$, presented above. In order to make sure that these results are representative, a careful analysis of the parametric space has been carried out (see S.I. for a full analysis). 
Profiles of the gap function considering different critical temperatures for $\omega=0.2$ eV are collected in S.I.
For a fixed critical temperature, it can be seen that the occupancies of the normal state deviate from the step function as the temperature increases, softening the transition around the Fermi energy. In all cases, the superconducting occupation numbers do not suffer big alterations with the temperature.  Hence, we will hereby take $T_c$=300 K for the analysis, and seize the effect of the changes in those functions with respect to the temperature, $T$. 

%In order to analyze the effects of superconductivity on a simple model of hydrogen-based superconductor, we will start with a tight binding model of a 1D hydrogen chain. The superconducting state will be modelled thanks to an approximation for the occupation numbers taken from SC-DFT (superconducting DFT)\cite{gross,gross2,Oliveira1988, Sanna17,sanna}, since it has been shown to very accurately reproduce experimental Tcs of conventional superconductors without introducing any empirical parameters.

%In order to apply these equations to the hydrogen chain tight-binding formalism, we have carried out a careful analysis of the parametric space, in order to make sure that it is representative of a physical system (see S.I. for a full analysis). 
%Noticeably, we took $a=2.64$ \AA\, in order to set the hydrogen-hydrogen distance to $d_{HH}=1.32$ \AA. 
%The Gaussian exponent was set to $\alpha=1.43$ \AA$^{-2}$ from fits.
%The hopping parameters, $v$ and $w$, were set from the H\"uckel model in the [-3.0,-1.5]$ eV$ range. We have chosen to set atomic-overlap dependant values, so that they reflect the variations across different interatomic distances.
%As far as the computational values are concerned, the real-space grid step, $\Delta x$, and the number of $k$ points were fixed from a convergence test to $\Delta x=10^{-3}\ a $ and $N_k=101$, respectively, ensuring errors in the orthogonality of the Bloch functions below $10^{-5}$.

\begin{figure}[h!]
    \centering
    \includegraphics[width=\linewidth]{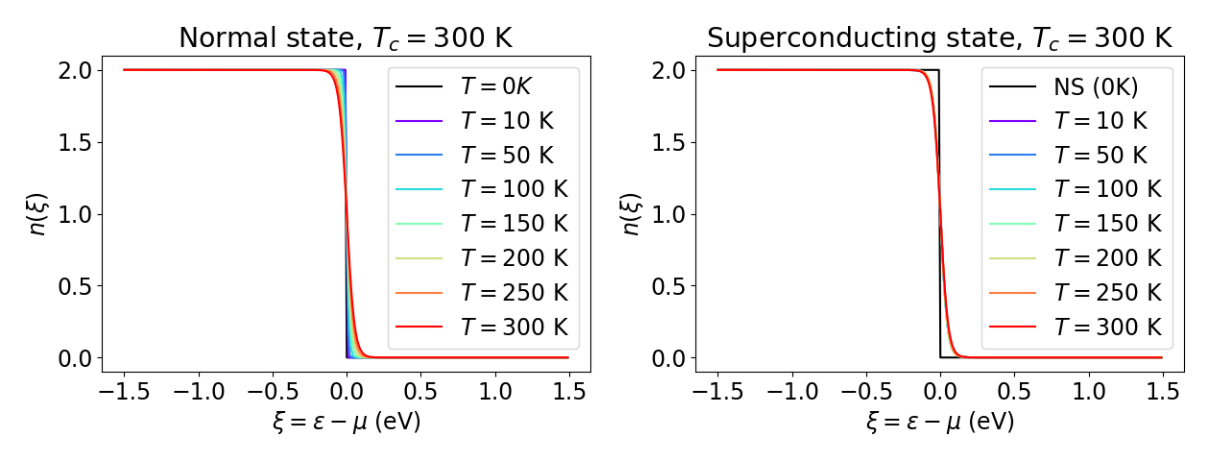}
    \caption{Occupation numbers at different energies. In all panels, that of the normal state at $T=0$ K is depicted in black. 
    To the left, the occupation numbers of the normal state at different temperatures. To the right, the same is displayed for the superconducting state with $T_c$=300 K.}
    \label{fig:occ-tc-t}
\end{figure}

\begin{figure}
    \centering
    \includegraphics[width=\linewidth]{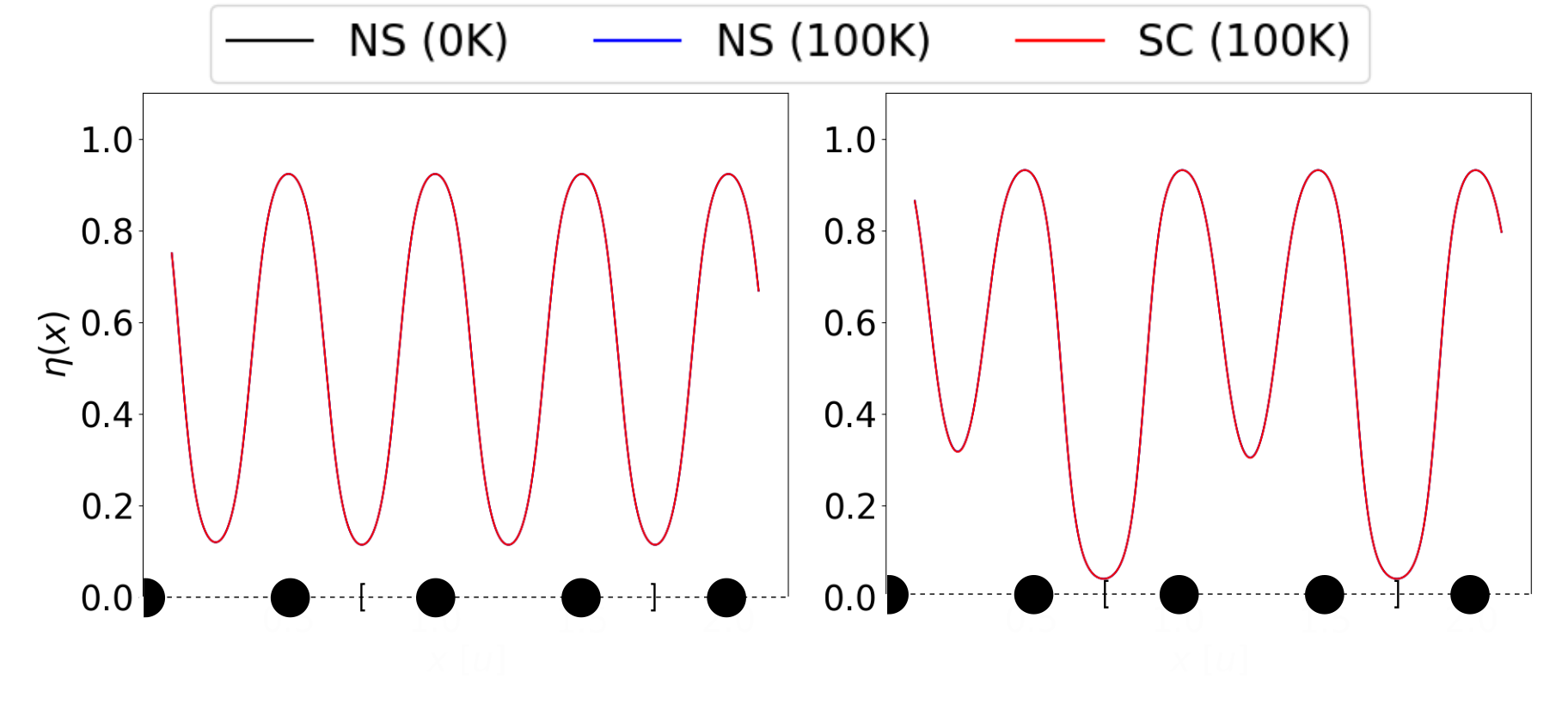}
    \caption{Profile of the ELF in the hydrogen chain for $T_c=300$ K. Left: metallic, right: dimerized. In both panels, the three overlapping lines represent the three considered models: normal state at $T=0$ K and $T=100$ K, and superconducting state at $T=100$ K. Atomic positions are marked by black circles, and the unit cell is delimited by squared brackets.}
    \label{fig:symmetric-elf}
\end{figure}

In the following, we denote the normal and superconducting states as NS and SC, respectively. Taking the results for the NS and SC occupation numbers into account it is not surprising that when we analyze the SC-ELF in the homogeneous hydrogen chain 
%(all interatomic distances and hoppings being the same and equal to $t=-1.5$ eV and $d_{HH} = 1.32\, $\AA) 
we see that it does not differ from the localization in the normal state (both T = 0 K and at T = 100 K are analyzed in Fig. \ref{fig:symmetric-elf}-left). This is true for the whole range of temperatures and interatomic distances.

Given that differences between the NS and SC seem to be very small, it becomes interesting to study the limiting case in which the superconducting properties are amplified, to assess the effect of the highly correlated state in the spatial properties.
The limit case of highest correlation in the superconducting state corresponds to the case when $\Delta\to \infty$ and $T\to 0$. The first condition can also be reformulated as $T_c\to \infty$. From eq.~\eqref{eq:scdft-rho-NS}, it is possible to infer that the occupancies will go to $1$ in those places where $\Delta(\xi)\neq 0$. Taking the very large value of $T_c=5000$ K and a low temperature of $T=10$ K, we obtain the occupation numbers shown in the inset of Figure 9
%{\color{red}XX} 
in SI.
Even at this limiting case the profile of the real-space functions in the symmetric chain in the SC state, namely the electron density, the KED, and the ELF; seem indistinguishable from those of the normal state. 

This result has big implications as far as the analysis of pairing in supercoductivity is concerned:  SC electron localization descriptors can be inferred from the normal state. Hence, with SC-ELF and the hydrogen chain enable us to rationalize the fact that the electron pairing can be obtained for superconductors at the DFT level from the normal state.

\begin{figure}[h]
    \centering
    \includegraphics[width=\linewidth]{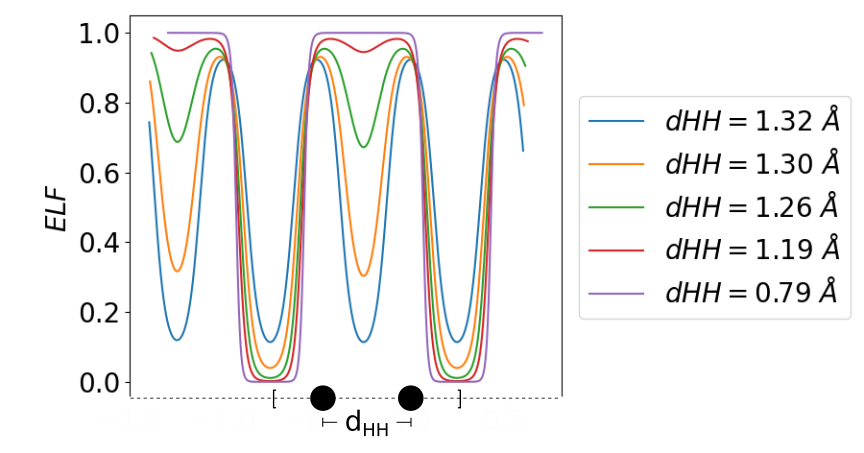}
    \caption{Normal state ELF profile for the dimerized hydrogen chain at $T=100$ K,
    %\textcolor{red}{remBB : maybe we shall add a phrase in the text to justify this value} 
    considering different minimum interatomic distances, $d_{HH}$. }
    \label{fig:ELF-T100-dim-all}
\end{figure}

%\subsection{Molecularity}
The 1D model also enables to understand the effect of bonding in metallization. 
When the distance between the two subcells is changed, dimerization is simulated (see Fig. \ref{fig:ELF-T100-dim-all}).
The profile of the ELF reveals the molecularization of the system as two of the hydrogens approach each other, with the appearance of two ELF local minima, a higher minimum between the two hydrogen forming a dimer and a lower minimum occurring between two unit cells (at mid-distance of two dimers).
The minimum value of the ELF, $\phi$, drops to nearly zero when slightly shifting the distances.
%from $d_{HH}=1.32$ \AA~to $d_{HH}=1.3$ \AA. 
Note that this ELF value would correspond to the networking value of the 1D-chain, as introduced by the authors in Ref.~\cite{Belli21}. Since this value was found to correlate with $T_c$, this would mean that the molecularity  hampers superconductivity. This result agrees with the proposal by Ackland et al from MD simulations \cite{molecularity0} and the high $T_c$ found for molecular systems in \cite{Belli21}.

Simultaneously, the value of the ELF at the higher local minima between the hydrogens in the same unit cell increases with respect to the symmetric case. This is characteristic of a more delocalized behavior of electrons in the intramolecular region. We shall call the value of the ELF at this local minima $\phi^*$, or Molecularity index, as it represents the first characterization of the molecularization effect.

Further decreasing the minimum interatomic distance accentuates these changes in the topology of the function.
In fact, when $d_{HH}$ decreases sufficiently, the atoms in the lattice form units resembling $H_2$ molecules, characterized by flat ELF profiles within the molecule. 
%We thus refer to this distance as intramolecular.

As in the Su-Schrieffer-Heeger (SSH) model, a gap opens upon dimerization. Hence, we have also plotted 
the evolution of the topological descriptors $\phi$ and $\phi^*$ with respect to the energy gap (see Fig. 10
%\ref{fig:phi-dHH-gap} 
in SI). It can be seen that a large value for $\phi^*$ is a feature of an insulating state in the hydrogen chain, where intramolecular distances are shortened. 
This is further supported by the increase of the localization in the ELF basins with the decreasing of the intramolecular distance, as it can be seen in Figure \ref{fig:ELF-T100-dim-all}.

%\section{Molecularity in real systems}
With this quantitative characterization of molecularity in the model at the DFT level, it is then possible to envisage its  characterization in real systems.  
We define the molecularity index in a 3D system, $\phi^*$, as the maximum value of the ELF function for which at least two hydrogen atoms become connected. In molecular systems, this will necessarily correspond to the value of the ELF for which molecular units appear, so that the number of atoms will be two. 
%A visual example is shown in S.I.

\begin{figure}
    \includegraphics[width=\linewidth]{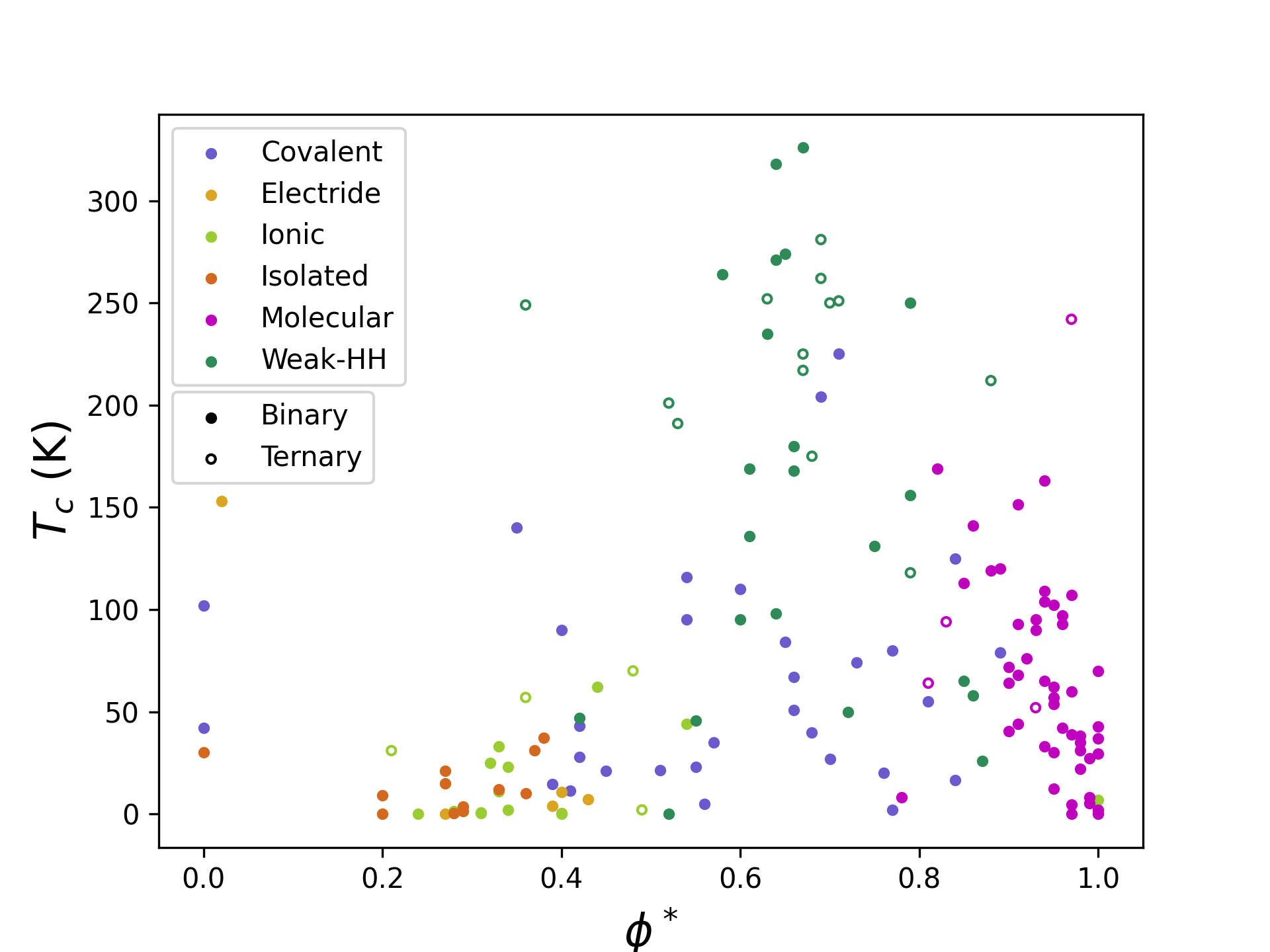} 
    \caption{Reference critical temperature $T_c$ (K) with respect to  the molecularity index $\phi^*$ for all binary and ternary data, classified by bonding type families. 
    }
    \label{fig:all-deltaphi.png}
\end{figure}

In order to test its use in complex systems, we have calculated it along with the networking value for a set of 129 binary and 21 ternary compounds (see S.I. for details). As expected, $\phi^*$ is high for molecular systems, ranging between $0.8$ and $1.0$, where no other type of bonding family is present (Fig. \ref{fig:all-deltaphi.png}). We notice that the two bond categories that are most likely to have a high $T_c$, namely covalent and weak H-H, are dominant in the region where $\phi^*\in [0.45, 0.8]$, meanwhile other types of bonding show generally low critical temperatures. 
In other words, the molecularity index is a quantitative tool to separate the families of interest. Thus, the molecularity index allows an automated characterization of bonding type, and hence of potential high $T_c$ superconductors. 

\begin{figure}
    \centering
    \includegraphics[width=\linewidth]{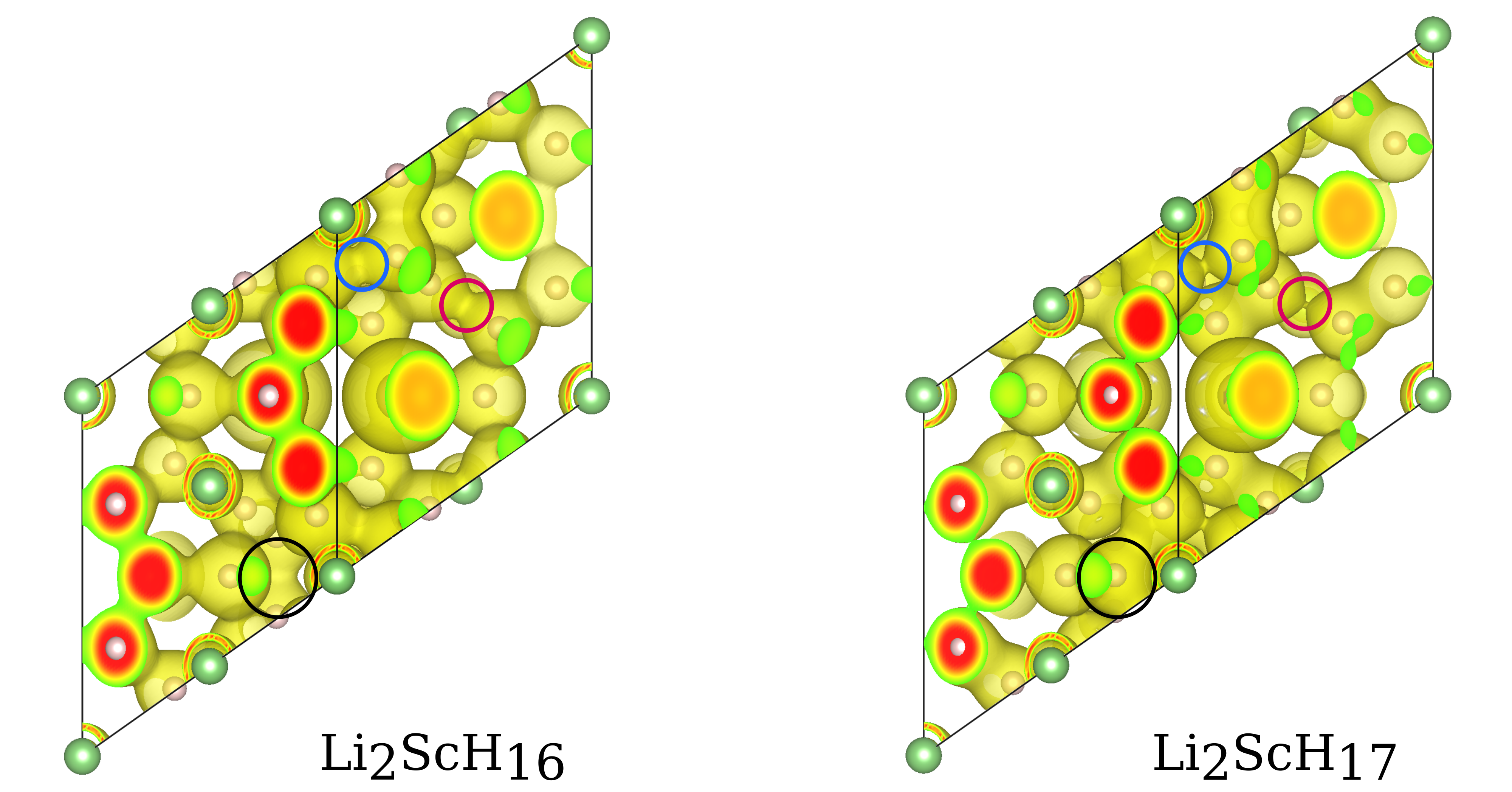}
    \caption{Isosurface of ELF$=0.57$ for Li$_2$ScH$_{16}$ (left, $\phi=0.63$) and Li$_2$ScH$_{17}$ (right, $\phi=0.57$) at $300$ GPa. The black circle marks the place where the extra H is added. The blue and red circles show how the values of the ELF at the critical points change with this addition, becoming more prone to form molecular units.}
    \label{fig:Li2ScHx}
\end{figure}

Moreover, a careful inspection of the new systems clearly shows that the molecularity index is necessary as the complexity of the systems increase, enabling to differentiate systems with similar networking value but very different $T_c$. 
This is the case of Li$_2$ScH$_{16}$ at 300 GPa, with $\phi=0.63$ and $T_c=281$ K, and Li$_2$ScH$_{17}$ at the same pressure, with $\phi=0.57$ and $T_c=94$ K.  
%The inclusion of $H_f$ and $H_{DOS}$ \textcolor{red}{remBB: what does this mean?...} does not seem to help, as they give $\Phi_{DOS} = 0.47$ and $\Phi_{DOS} = 0.41$, respectively. 
The severe drop of the critical temperature upon inclusion of one extra hydrogen atom is understood when one examines the ELF isosurfaces: the hydrogen atoms rearrange to shorten the minimum distance between them, and as a result they form molecular units  (see Fig.~\ref{fig:Li2ScHx}). This can be measured by an increase of the molecularity index from $\phi^*=0.69$ in Li$_2$ScH$_{16}$ to $\phi^*=0.83$ in Li$_2$ScH$_{17}$, the latter being above the threshold of $\phi^*=0.8$ for molecular systems. As proved in the 1D chain, this is detrimental for high-temperature superconductivity \cite{Belli21, Saha2023}.
Hence the new index complements the networking value when going to complex systems, enabling to characterize  complex ternary superconductors.

\begin{figure}[h!]
    \centering
    \includegraphics[width=\linewidth]{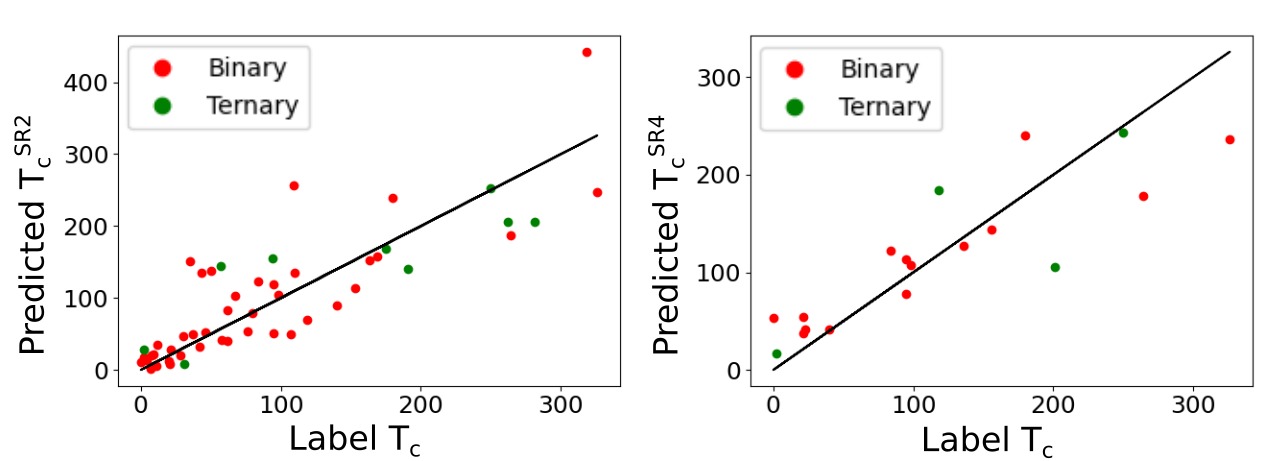}
    \caption{Predicted values of $T_c$ (K) with respect to the reference (label) data, as computes using the fits SR2 (left) and SR4 (right). For the latter, only systems where $\phi^* \in [0.45, 0.8]$ are considered.}
    \label{fig:SRmolecularity}
\end{figure}

With the new derived index, it becomes then necessary to propose a new expression to calculate $T_c$ that works both for binary and ternary compounds. Resorting to symbolic regression (see S.I. for details), it is possible to propose a new fit which leads to better accuracy and wider applicability. We propose the following expressions: %$T_c^{SR1}=418.37(1-\Delta\phi)\sqrt{H_{DOS}}H_f^3$
\begin{eqnarray}
\label{eq:SR1}
T_c^{SR1} &=& 382.5\, (1-\Delta\phi) H_f H_{DOS}\, ,\\    
T_c^{SR2} &=& 442.3\, (1-\Delta\phi) H_f^3 \sqrt{H_{DOS}}\, ,
\label{eq:SR2}
\end{eqnarray}
where $\Delta\phi=\phi^*-\phi$ takes into account both the networking and the molecularity indexes,  $H_{DOS}$ is the fraction of the density of states (DOS) at the Fermi level that correspond to Hydrogen and $H_f$ is the fraction of atoms in the unit cell.
Indeed, this expression is able to differentiate between ternary compounds with similar stoichiometries and different $T_c$'s, leading to MAEs of 38 K and 36 K for eqs.~\ref{eq:SR1} and \ref{eq:SR2}, respectively, in a set of systems that was not used to fit those expressions. Those equations should be compared with the expression from Ref.~\cite{Belli21}: $T_c = (750  \phi \, H_f \sqrt[3]{H_{DOS}} - 85)K$; with MAE of 55 K in the same dataset. %Using eqs.~\ref{eq:SR1} and \ref{eq:SR2}, the MAE in the training set is of 35 K and 32 K for models SR1 and SR2, respectively.
The comparison between the predictions in the test set for SR2 are displayed in Fig.~\ref{fig:SRmolecularity}-left. Further improvements can be even achieved from the observation that the estimates also improve in the high $T_c$ region, where predictions are harder due to the little availability of data (see S.I.). 

This improvement is due to the fact that $\phi^*$ is able to identify bonding types. Hence, we can also use this ability to filter the data so as to keep  the systems with $\phi^* \in [0.45, 0.8]$, which correspond to the bonding families that interest us. (see S.I. for details).
Using these data, a much sharper correlation is observed (see Fig. 13 of S.I.). Several analytical expressions are proposed, where the overall errors are much more consistent with those in high-$T_c$ regions, thus being more reliable for our purposes:
%\textcolor{red}{change here by the expressions with molecularity included-also in SI}
\begin{eqnarray}
%    T_c^{SR2} &=& 249 H_f (\phi^2+H_{DOS}),\\
%    T_c^{SR3} &=& 784 \phi^3 H_f^3 + 175 H_{DOS},\\
%    T_c^{SR4} &=& 625 \phi^3 H_f^2 + 219 H_f H_{DOS}
    T_c^{SR3} &=& 312.0 \, H_{DOS},\\
    T_c^{SR4} &=& 574.7 \, \phi \sqrt{H_f^3\, H_{DOS}}.
\end{eqnarray}
The results of the predicted $T_c$ in the test set as obtained with the fit SR4 can be visualized in Fig.~\ref{fig:SRmolecularity}-right. For high-throughput analysis, all four models are recommended in order to mutually discard outliers. We believe these new expressions should help in a better prediction and high-throughput analysis of potential high Tc hydrogen based superconductors.

%\section{Conclusions}

All in all, we have resorted to the SC-DFT approach to develop a new formulation of superconducting ELF in terms of a reorganization of occupation numbers with respect to the normal state. We have first applied these new developments to a model system, leading to two main conclusions. On the one hand, the small changes in occupation numbers lead to small changes in localization from the normal to the superconducting state. 
Taking into account that most calculations in solid state are carried out within the DFT framework, where 2-body quantities are not usually accessible, having a DFT-based index that only requires Kohn-Sham orbital information  enables for a quick screening of the chemistry in potential superconductors. 
On the other hand, the evolution of ELF upon changes in dimerization prognoses a lowering of the critical temperature upon formation of H$_2$ molecules. Moreover, the simple picture in the 1D-chain enables to introduce a molecularity index that allows to quantify this process. 
Building from these results, we have applied the molecularity index to 3D systems and calculated it from a set of binary and ternary systems, showing that i) it allows for the first time to automatically classify  the bonding type of these systems and ii) it allows to differentiate tricky situations where molecules appear and previous indexes, developed for binary systems, fail. Hence, this new index goes beyond current proposals (which fail for ternary compounds), allowing for the automatic characterization of complex potential superconducting  compounds in a fast manner, with especial emphasis on high $T_c$ prediction. These theoretical advances will help pushing the inverse design of high T$_c$ superconductors to a reliable and cost-efficient limit.

%\section{Acknowledgements}
We would like to acknowledge support by ECOS-Sud C17E09 and C21E06, and the Association Nationale de la Recherche under grant ANR-22-CE50-0014. This research was supported by the European Research Council (ERC) under the European Unions Horizon 2020 research and innovation programme (grant agreement No. 810367), project EMC2.
I.E. acknowledges funding from ERC under the European Unions Horizon 2020 research and innovation program (Grant Agreements No. 802533 and No. 946629); 
the Department of Education, Universities and Research of the Eusko Jaurlaritza, and the University of the Basque Country UPV/EHU (Grant No. IT1527-22); and the Spanish Ministerio de Ciencia e Innovación (Grant No. PID2022- 142861NA-I00). 

\bibliography{main}

\end{document}

\title{\vspace*{-.55cm} \Large \bf Supplementary Information}

\date{\today}

\maketitle
\section{Superconducting Density Functional Theory}
\label{section:SCDFT}

The widespread use of DFT for electronic structure calculations, due to its great compromise between accuracy and computational time, has served as a motivation to extend it to a wider variety of systems. The case of superconductors is particular in the sense that it cannot be solved in a perturbative fashion. In fact, this is so because in these systems the phase symmetry is broken, which implies that the number of particles will not be conserved. Another key property of superconductors is that the  coupling between electrons and phonons becomes important, and that it is in fact this what permits the superconductivity. The ionic displacements around their equilibrium positions must therefore be taken into account in the Hamiltonian. This is why, in Superconducting Density Functional Theory (SCDFT) \cite{gross,gross2,Oliveira1988,Sanna17,sanna}, we must consider the nuclear degrees of freedom.
%, as in equation \eqref{eq:molec-H}. 
The purely ionic part is written in terms of the ionic field operators, $\Phi(\bm{R})$, and contains the kinetic and interaction terms,
\begin{eqnarray}
\notag 
&\hat H_i& = -\int \Phi^\dag(\bm{R})\frac{\nabla^2}{2M} \Phi(\bm{R}) d\bm{R} \\ &+& \frac{1}{2} \int  \Phi^\dag(\bm{R})\Phi^\dag(\bm{R}') \frac{Z}{\abs{\bm{R}-\bm{R}'}} \Phi(\bm{R}')\Phi(\bm{R})\, d\bm{R}\, d\bm{R}'\, .
\end{eqnarray}
Meanwhile, the interaction between electrons and ions is
\begin{equation}
\hat H_{ie} = -\frac{1}{2} \sum_{\sigma}\int \psi^\dag_{\sigma}(\bm{r})\Phi^\dag(\bm{R})\frac{Z}{\abs{\bm{R}-\bm{r}}} \Phi(\bm{R})\psi_{\sigma}(\bm{r})\,d\bm{R} \, d\bm{r} \,  .
\end{equation}
For the electronic part the Hamiltonian takes the form, 
\begin{eqnarray}
\notag 
&\hat H_e& = \sum_{\sigma} \int \psi^\dag_{\sigma}(\bm{r})\left[-\frac{\nabla^2}{2} - \mu\right] \psi_{\sigma}(\bm{r})\, d\bm{r} \\ &+& \frac{1}{2} \sum_{\sigma\sigma'} \int \psi^\dag_{\sigma}(\bm{r})\psi^\dag_{\sigma'}(\bm{r}') \frac{1}{\abs{\bm{r}-\bm{r}'}} \psi_{\sigma}(\bm{r})\psi_{\sigma'}(\bm{r}')\, d\bm{r}\, d\bm{r}' \, .
\end{eqnarray}

Finally, we must consider three different external potentials: one that couples to electrons, $v_{ext}(r)$, one that couples to ions, $W_{ext}({\bm{R}})$, and an anomalous potential, $\Delta_{ext}(r,r')$, that is responsible for the symmetry breaking, allowing Cooper pairs to tunnel in and out of the system,
\begin{equation}
\hat H_{\Delta_{ext}} = \int \Delta^*_{ext}(\bm{r},\bm{r}') \psi_{\uparrow}(\bm{r})\psi_{\downarrow}(\bm{r}')\, d\bm{r} \,d\bm{r}' + h.c. 
\end{equation} 
If we let this quantity go to zero, the Hamiltonian converges to a non-superconducting one, i.e. that of a normal state system. In this way, SCDFT  allows to \emph{turn on and off} superconductivity, and to compare the system's properties when it is in the normal or in the superconducting state.

As in usual DFTs, there will be a Hohenberg-Kohn theorem establishing a one-to-one mapping between, in this case, those three external potentials and corresponding densities. Within this approach the Hamiltonian includes a non-particle conserving superconducting symmetry breaking term and the theory is formulated in the grand-canonical ensemble.  The expectation values are thus taken considering the grand-canonical density matrix,
\begin{equation}
\hat \rho_0=\frac{e^{-\beta(\hat H-\mu \hat N)}}{Tr[e^{-\beta(\hat H-\mu \hat N)}]}\, ,
\end{equation}
where $\hat N$ is the number operator. Then, the electron density is
\begin{equation}
    \rho^{SC}(\bm{r}) = \left\langle \sum_{\sigma} \psi^\dag_{\sigma}(\bm{r})\psi_{\sigma}(\bm{r})\right\rangle = Tr\left[\hat \rho_0 \sum_{\sigma}\psi^\dag_{\sigma}(\bm{r}) \psi_{\sigma}(\bm{r})\right]\, ,
    \label{eq:scdft-rhosc-1}
\end{equation}
where we have used the \emph{SC} superscript to differentiate it from the normal state electron density. 

In a similar way, we define the ionic density,
\begin{equation}
    \Gamma(\{\bm{R}_i\}) = Tr\left[\hat \rho_0 \prod_{j}\Phi^\dag(\bm{R}_j)\Phi(\bm{R}_j)\right]\, .
    \label{eq:scdft-Gamma}
\end{equation}
Just as did the ionic external potential, this density depends on \emph{all} ionic coordinates, thus corresponding to an N-particle density matrix. This will be convenient when we start treating the ionic deviations from equilibrium positions as collective vibrations (phonons). 
Finally, we define the anomalous density, 
\begin{equation}
    \chi(\bm{r}, \bm{r}') = Tr\left[\hat \rho_0\,  \psi_{\uparrow}(\bm{r})\psi_{\downarrow}(\bm{r}')\right]\, ,
    \label{eq:scdft-chi-1}
\end{equation}
which is a two-body object and the order parameter of the transition. 
%Notice that $\chi(r,r')$ is analogue to the singlet-pair function of equation \eqref{eq:SP-wfn-1}. 

A second Hohenberg-Kohn-like theorem proves that the grand-canonical potential follows a variational principle as a function of the densities in \eqref{eq:scdft-rhosc-1}, \eqref{eq:scdft-Gamma} and \eqref{eq:scdft-chi-1}. This potential is defined as
\begin{eqnarray}
\notag 
&\Omega&\left[\rho^{SC}, \chi, \Gamma \right] = T_e\left[\rho^{SC}, \chi, \Gamma \right] + T_n\left[\rho^{SC}, \chi, \Gamma \right] \\
\notag
&-& \frac{1}{\beta}S\left[\rho^{SC}, \chi, \Gamma \right] + V_{ee}\left[\rho^{SC}, \chi, \Gamma \right] + V_{nn}\left[\rho^{SC}, \chi, \Gamma \right] \\
\notag  
&+& \int v_{ext}(\bm{r})\rho^{SC}(\bm{r})\, d\bm{r} + \int W_{ext}(\{\bm{R}_i\})\Gamma(\{\bm{R}_i\})\prod_j d\bm{R}_j \\ 
&+& \int \Delta^*_{ext}(\bm{r}, \bm{r}') \chi(\bm{r}, \bm{r}')\, d\bm{r} \,d\bm{r}' \\
\notag &\equiv& F\left[\rho^{SC}, \chi, \Gamma \right] +  \int v_{ext}(\bm{r})\rho^{SC}(\bm{r})\, d\bm{r} \\ 
\notag  
&+& \int W_{ext}(\{\bm{R}_i\})\Gamma(\{\bm{R}_i\})\prod_j d\bm{R}_j \\
 &+& \int \Delta^*_{ext}(\bm{r}, \bm{r}') \chi(\bm{r}, \bm{r}')\, d\bm{r} \,d\bm{r}'\, ,
\end{eqnarray}
where $T$ corresponds to kinetic energy functionals and $S$ to the entropy. We have also introduced the system-independent universal functional, $F\left[\rho, \chi, \Gamma \right]$, analogue to the DFT functional.%of eq.~\eqref{eq:univ-func}. 
Then, the variational principle ensures $\Omega[\rho^{SC}_0,\chi_0,\Gamma_0]<\Omega[\rho^{SC},\chi,\Gamma]$ for any $\rho^{SC}$, $\chi$, $\Gamma$ different from the ground state ones, $\rho^{SC}_0$, $\chi_0$, $\Gamma_0$. 
%The proof can be found elsewhere \cite{Mermin1965}.

As usual, we do not know how $F[\rho^{SC},\chi,\Gamma]$ looks in terms of the densities, so we will have to introduce a reference (Kohn-Sham) system with non-interacting electrons whose grand-canonical potential is minimized by the same densities. The resulting equation for the ions is analogue to Kohn-Sham, but considering the corresponding ionic potential.
%\begin{align}
%    \left[\sum_{A} \frac{\nabla^2_A}{2M_A} + W_{s}(\{\RR_A\})\right]\Phi_n(\{\RR_A\}) = \mathscr{E}_{n} \Phi_n(\{\RR_A\})\, .
%\end{align}
Meanwhile, the Kohn-Sham electronic equations yield a Hamiltonian of the form
\begin{eqnarray}
\notag
    \hat{H}_s &=& \sum_{\sigma} \int d\bm{r}\, \psi^\dag_{\sigma}(\bm{r}) \left[-\frac{\nabla^2}{2} + v_{KS}(\bm{r}) - \mu\right] \psi_{\sigma}(\bm{r}) \\ 
    &+& \int d\bm{r} d\bm{r}' \left[\Delta_{KS}^* (\bm{r},\bm{r}')\psi_{\uparrow}(\bm{r})\psi_{\downarrow}(\bm{r}') + h.c. \right]\,  ,
    \label{eq:scdft-elec-H}
\end{eqnarray}
where $v_{KS}(r)$ is the usual electronic KS potential.
%(see eq.~\eqref{eq:v_KS}).
%and the effective anomalous potential takes the form %the anomalous potential $\Delta_s^* (\R,\R')$ couples the system to an external source of Cooper pairs. 
%\begin{align}
%\Delta_{KS} (r,r') = \Delta_{ext}(r,r') + \frac{\delta F_{xc}[\rho^{SC}, \chi,\Gamma]}{\delta \chi}\, ,
%\end{align}

\section{The Electron Localization Function (ELF)}

The electron localization function was first introduced by Edgecome and Becke to identify regions of localized same-spin electron pairs, or groups of them, in atomic and molecular systems.\cite{Becke1990} It is based on the same-spin pair probability as approximated in Hartree-Fock,
\begin{align}
P^{\sigma\sigma}_2 (\bm{r}_1,\bm{r}_2) = \rho_{\sigma}(\bm{r}_1)\rho_{\sigma}(\bm{r}_2) - \abs{\rho^{\sigma}_1(\bm{r}_1,\bm{r}_2)}^2 \, ,
\label{eq:prob12-hf}
\end{align}
which is the probability of finding simultaneously one electron with spin $\sigma$ at $\bm{r}_1$, and another $\sigma$ electron at $\bm{r}_2$. Here, $\rho^{\sigma}_1(\bm{r}_1,\bm{r}_2)$ is the spin-resolved 1-RDM, and $\rho^{\sigma}(\bm{r}_1)$ its diagonal, which corresponds to the electron density for $\sigma$-spin. In eq.~\eqref{eq:prob12-hf}, we recognize the first term as the product of the individual probabilities of finding each electron at the said positions, to which we need to substract a term arising from the correlation introduced in Hartree-Fock by the Pauli principle (exchange interaction).

If we assume that there is one $\sigma$ electron at $\bm{r}_1$, we can express the probability of finding another electron with the same spin at $\bm{r}_2$ by
\begin{align}
    P^{\sigma\sigma}_{cond} (\bm{r}_1,\bm{r}_2) &=  \frac{P^{\sigma\sigma}_2 (\bm{r}_1,\bm{r}_2)}{\rho_{\sigma}(\bm{r}_1)} = \rho_{\sigma}(\bm{r}_2) - \frac{\abs{\rho^{\sigma}_1(\bm{r}_1,\bm{r}_2)}^2}{\rho_{\sigma}(\bm{r}_1)} \, ,
\end{align}
which we call the conditional same-spin pair probability. Fixing one of the electrons lets us study the behavior of this probability when $r_2 \to r_1$. Changing to the spherically averaged version of $P^{\sigma\sigma}_{cond}$, that depends on the coordinates $(\bm{r}, \bm{s})$, where $\bm{r}$ is the reference point and $s$ a distance from it, and doing a Taylor expansion, we can show that
\begin{align}
    P^{\sigma\sigma}_{cond}(\bm{r},s) = \frac{1}{3} \left[\sum_{i}^{\sigma}\abs{\nabla\phi_i(\bm{r})}^2 - \frac{1}{4}\frac{\abs{\nabla\rho_{\sigma}(\bm{r})}}{\rho_{\sigma}(\bm{r})} \right] s^2 + \dots
\label{eq:prob-cond-sph}
\end{align}
Here, $\phi_i(\bm{r})$ are the HF orbitals, and the sum $\sum_{i}^{\sigma}$ means that we are only considering the orbitals containing electrons of spin $\sigma$.

From \eqref{eq:prob-cond-sph}, we recognize the term in brackets,
\begin{align}
    D_{\sigma}(\bm{r}) = \sum_{i}^{\sigma}\abs{\nabla\phi_i(\bm{r})}^2 - \frac{1}{4}\frac{\abs{\nabla\rho_{\sigma}(\bm{r})}}{\rho_{\sigma}(\bm{r})}\, ,
    \label{eq:Dr}
\end{align}
as a measure of localization, as it is the leading term for small distances $s$ between the electrons. This is, when $D_{\sigma}$ is small, the probability of finding a $\sigma$ electron very close to the reference one is also small. This means that the reference electron is very localized, and so is the Fermi hole that comes with it, not allowing a same-spin electron to come near. The opposite is true when the reference electron is delocalized: it is more likely that the other electron comes close to the reference position, and $D_{\sigma}$ will be high. 

All of this has been introduced for electrons of the same spin. However, if the reference electron is very localized, it is likely than an opposite electron is also very localized in that region, as it will also be avoiding its own same-spin electron, thus forming a singlet pair. This is a consequence of the exchange introduced in the same-spin probability in eq.~\eqref{eq:prob12-hf}.

The function $D_{\sigma}(\bm{r})$ being opposite to localization, we introduce the Electron Localization Function (ELF),
\begin{align}
    \eta(\bm{r}) = \left[1+\left(\frac{D_{\sigma}(\bm{r})}{D^0_{\sigma}(\bm{r})}\right)^2\right]^{-1}\, ,
    \label{eq:elf-spin}
\end{align}
where 
\begin{align}
    D^0_{\sigma}(\bm{r}) = \frac{3}{5} (6\pi^2)^{2/3} \rho_{\sigma} (\bm{r})^{5/3}
\end{align}
is the term $D_{\sigma}$ as evaluated for a uniform electron gas. This normalization allows to compare the values of the kernel $\chi_{\sigma}(\bm{r})=D_{\sigma}(\bm{r})/D^0_{\sigma}(\bm{r})$ of different systems. Further, the Lorentzian transformation applied on that kernel in the definition of $\eta(\bm{r})$ in \eqref{eq:elf-spin} allows us to have a function that ranges between 0 and 1, that is high in the regions of high localization ($\eta\to 1$), and that conserves the topology of $\chi_{\sigma}(\bm{r})$. This, as we shall see, will be very important in the description of electron localization.

Although the ELF was first introduced in the context of the HF approximation, a close inspection of eq.~\eqref{eq:Dr} allows us to identify the first and second terms as the kinetic energy density of the system, $\tau(\bm{r})$, and its form in the von-Weizsacker approximation, $\tau_{vW}(\bm{r})$ \cite{vonWeizsacker1935}. In fact, in a closhed-shell system we have $\rho(\bm{r})=2\rho_{\sigma}(\bm{r})$, and therefore we can define the spinless quantity
\begin{align}
    D(\bm{r}) &= \frac{1}{2} \sum_i \abs{\nabla\phi_i(\bm{r})}^2 - \frac{1}{8} \frac{\abs{\nabla\rho(\bm{r})}^2}{\rho(\bm{r})}\, ,\\
    &= \tau(\bm{r}) - \tau_{vW}(\bm{r})\, ,
\end{align}
and 
\begin{align}
    \eta(r) = \left[1+ \left(\frac{D(\bm{r})}{D_0(\bm{r})}\right)^2\right]^{-1}\, ,
\end{align}
with $D_0(\bm{r})=\frac{3}{10}(3\pi^2)^{2/3}\rho(\bm{r})^{5/3}$. This form of the ELF written in terms of kinetic energy densities was first introduced by Savin \cite{SavinELF}, and allows to compute it beyond the HF approximation. Further, it introduces a new interpretation: because the von-Weizs\"acker kinetic energy is exact for a bosonic system of the same density $\rho(\bm{r})$, the term $\tau(\bm{r})-\tau_{vW}(\bm{r})$ is a local measure of the excess kinetic energy due to the fermionic nature of the electrons, or what we call the \emph{Pauli kinetic energy}. If this is high, it means that electron pairs are delocalized in that region, and the ELF will be small. If the kinetic energy density is not locally increased as an effect of the exclusion principle, in that case we say that electrons are localized, which will be reflected on a high value of the ELF.

%\subsection{Bond classification}
The Electron Localization can be used to classify bonds in supercoductors along with charge analysis. Following Ref.~\onlinecite{Belli21}, we pinpoint six distinct families: molecular systems, covalent systems, systems influenced by weak covalent hydrogen-hydrogen interactions, systems exhibiting electride behavior, ionic systems, and isolated systems. In each instance, the bond nature is identified through analyzing ELF saddle points between different atoms.

\section{The tight-binding model in real space}
 
\begin{figure}
    \centering
    \includegraphics[width=0.6\linewidth]{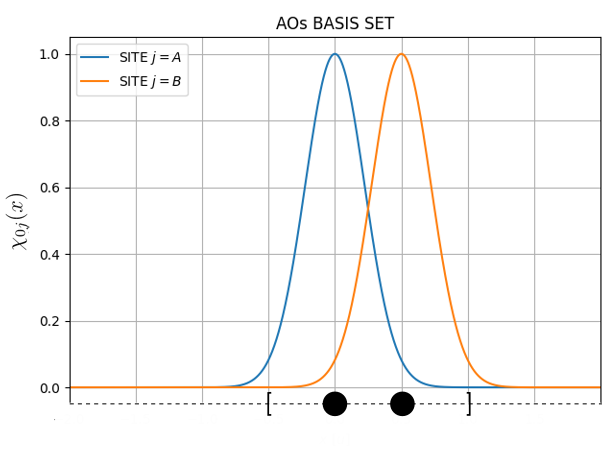}
    \caption{Atomic orbitals used in the unit cell of the hydrogen chain tight-binding model, represented by two identical Gaussian functions. The black spheres mark the positions of the two nuclei inside the unit cell, delimited by squared brackets. Here, the unit cell length is taken as $1\, u$, and the Gaussian exponent as $\alpha= 10\, u^{-2}$}
    \label{fig:AOs-tb}
\end{figure}

In order to make use of the tight-binding formalism to evaluate the real-space properties of the hydrogen chain, we begin by building a basis set of Gaussian atomic functions:
\begin{align}
    \chi_{lA}(x) &= e^{-\alpha(x-la)^2}\, ,\\
    \chi_{lB}(x) &= e^{-\alpha(x-(R_{AB}+la))^2}\, ,
\end{align}
that depends on the unit cell index, $l$. In this way, every unit cell contains two basis functions. Fig.~\ref{fig:AOs-tb} is a spatial representation of the two atomic orbitals in the unit cell. In general, we consider the length of the unit cell, $a$, to be equal to $1.0\, u$, with $u$ some arbitrary units. In the next section, we shall see which values of $a$ properly represent a hydrogen chain. The same is done for the Gaussian exponent $\alpha$.

\begin{figure}
    \centering
    \includegraphics[width=\linewidth]{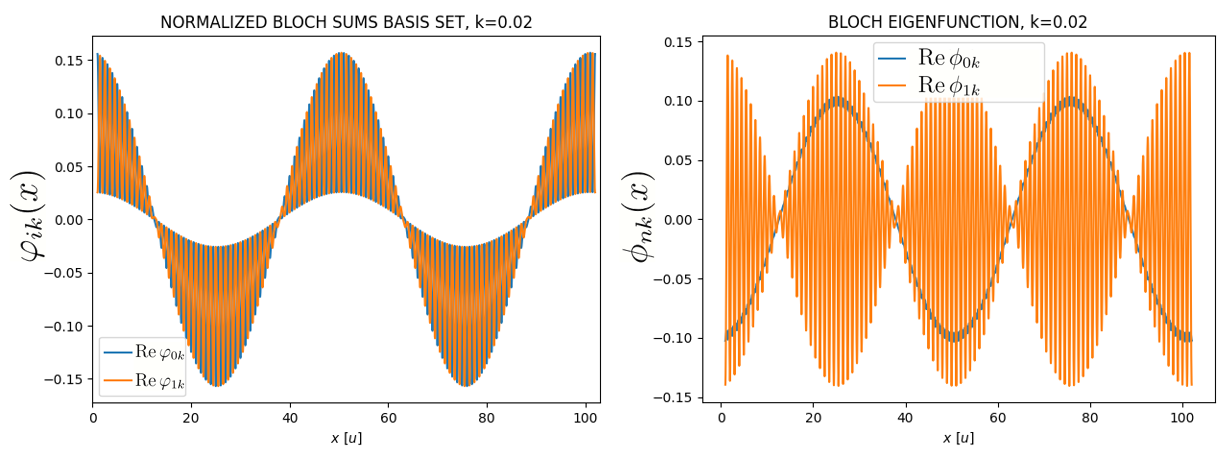}
    \caption{Left: Real part of the Bloch sums of each band index, $i=0$ and $i=1$, for $k=0.02\times\frac{2\pi}{a}$. Right: Real part of the Bloch wavefunctions at the same $k$ point for the symmetric chain.}
    \label{fig:bloch-basis-wfn}
\end{figure}

With this, the atomic orbitals fulfilling Bloch's theorem, or Bloch sums, are 
\begin{align}
    \label{eq:bloch-sum1}
    \varphi_{Ak}(x) &= \frac{1}{\sqrt{N_l}} \sum_{l} e^{ikla} e^{-\alpha(x-la)^2}\, ,\\
    \varphi_{Bk}(x) &= \frac{1}{\sqrt{N_l}} \sum_{l} e^{ik(R_{AB}+la)} e^{-\alpha(x-(R_{AB}+la))^2}\, ,
    \label{eq:bloch-sum2}
\end{align}
with $N_l$ the number of unit cells considered in the model. Here, $k$ is selected to take values inside the first BZ, $k \in\left[-\frac{\pi}{a},\frac{\pi}{a}\right]$, with $k=\frac{2\pi}{N_l a}l$ and $l$ taking integer values in the range $\left[-\frac{N_l}{2},\frac{N_l}{2}\right]$. For simplicity, we will generally refer to $k$ in units of $\frac{2\pi}{a}$. A graphical representation of these orbitals for $k=0.02$ (i.e. $k=0.02\times\frac{2\pi}{a}$) is offered in Figure \ref{fig:bloch-basis-wfn}, for a total number of $N_l=101$.

Finally, the Bloch wavefunctions that diagonalize the Hamiltonian of the hydrogen chain can be written as a linear combination of the Bloch sums in \eqref{eq:bloch-sum1} and \eqref{eq:bloch-sum2},
%, as in eq.~\ref{eq:bloch-wfns-tb},
\begin{align}
    \phi_{nk}(x) &= \frac{1}{\sqrt{N_l}} \sum_i\sum_{R_i}c_{ik}^n e^{ikR_i} e^{-\alpha(x-R_i)^2}\, .
\end{align}
In this case, the band index $n$ can take the values $0$ and $1$. The diagonalization determines the coefficients $c_{ik}^n$, ant it is performed using the Gaussian elimination method \cite{Riley2006}, as implemented in NumPy \cite{Harris2020}. The result of the diagonalization will necessarily depend on the hopping parameters. Let us recall that the hopping parameter between site $A$ and the site $B$ to its right is $v$, whereas between the same site $A$ and the atom $B$ on its left, the hopping parameter is $w$. In the special case in which $v=w$, the resulting coefficients are $c_{Ak}^0=c_{Bk}^0=\frac{1}{\sqrt{2}}$ and $c_{Ak}^1=-c_{Bk}^1=\frac{1}{\sqrt{2}}$, corresponding to a bonding and antibonding orbitals, respectively. Figure \ref{fig:bloch-basis-wfn} shows an example of such a Bloch wavefunction.

Once the Bloch wavefunctions have been defined and computed, the next step is to evaluate the real-space functions of interest. In order to do so, we will generally assume that the Bloch wavefunctions are orthonormal, which allows expressions such as 
\begin{align}
    \rho(x) = \sum_{nk} n_{nk} \abs{\phi_{nk}(x)}^2\, ,
    \label{eq:dens-tb}
\end{align}
for the electron density, and similar ones for the other functions. 

To evaluate the orthogonality, we first resort to the nearest-neighbors approximation, in which the only non-zero overlaps between the basis functions are
\begin{align}
    \int \chi^*_{lA}(x) \chi_{lA}(x) dx = \int \chi^*_{lB}(x) \chi_{lB}(x) dx = \mathcal{C}\, ,
    \label{eq:ao-norm}
\end{align}
and that between two adjacent atomic sites,
\begin{eqnarray}
\notag
    \int \chi^*_{lA}(x) \chi_{lB}(x)\, dx  &=& \int \chi^*_{l-1B}(x) \chi_{lA}(x)\, dx \\ 
    &=& \int \chi^*_{lB}(x) \chi_{l+1A}(x)\, dx = \mathcal{S}\, 
    \label{eq:overlap-basis}
\end{eqnarray}
because we consider PBC, we must also include
\begin{align}
    \int \chi^*_{N_lB}(x) \chi_{1A}(x)\, dx = \mathcal{S}\, .
\end{align}
Equation \eqref{eq:ao-norm} allows to normalize the basis set. Once this is done, the Bloch overlap will yield
\begin{align}
    \int\varphi^*_{ik}(x)\varphi_{jk'}(x)\, dx = \delta_{k,k'}c_{i,j}^{k,k'}\, ,
    \label{eq:overlap-sums}
\end{align}
where it can be shown that $c_{A,A}^{k,k}=c_{B,B}^{k,k}=1$, and $c^{k,k}_{i,j}$ for $i\neq j$ is dependent on $\mathcal{S}$. 
In this way, equation \eqref{eq:overlap-sums} shows that the Bloch sums are orthogonal with respect to the wave vector index k, but not with respect to the site. This changes when we diagonalize the Hamiltonian and find the Bloch eigenfunctions, which are orthonormal
\begin{align}
    \int \phi^*_{nk}(x) \phi_{n'k'}(x) \, dx = \delta_{n,n'}\delta_{k,k'}\, .
    \label{eq:ortho-bloch-wfn}
\end{align}

Equation \eqref{eq:ortho-bloch-wfn} is true analytically, but in practice there is a numerical error that comes from the integration of the Bloch sums in \eqref{eq:overlap-sums}: the overlap of two Bloch sums of different $k$ is not exactly zero. Given the importance of the orthogonality condition in the computation of the density and the other real-space descriptors, we define the numerical error associated to those computations as
\begin{align}
    ERROR = \max_{k\neq k'} \left\{ \int\varphi^*_{ik}(x)\varphi_{jk'}(x)\, dx \right\}\, .
    \label{eq:ERROR-overlap}
\end{align}
This quantity sets the precision of the results presented in the following, and it will be of special interest to minimize it when choosing the parameters of the model.

As a final remark concerning the computation of the real-space quantities in this model, it is important to note that the expression of the ELF is not valid for one-dimensional systems, as the expression of the TF KED changes in those cases to \cite{Trappe2023}
\begin{align}
    \tau_{TF} (x) = \frac{\pi^2}{24}\, \rho(x)^3\, .
\end{align}
It is this one-dimensional version of the TF KED and the corresponding ELF, that we consider for all of the calculations of this Section.

\section{The model parameters}

The construction of the hydrogen chain using the tight-binding formalism relies on a set of parameters that need to be chosen carefully for the model to properly represent the physical system. Firstly, we choose the lattice parameter, $a$, for which we hereby adopt the value of $a=2.64\,$\AA.$\,$ This determines the value of the arbitrary units, $1\, u= 2.64\,$ \AA. This value is in accordance with what has been observed in the literature for high-temperature hydrogen-based superconductors of interest \cite{Belli21, Saha2023}. In this way, for the symmetric chain we shall set the hydrogen-hydrogen distance to $d_{HH}=0.5\, u = 1.32\,$ \AA.

The Gaussian exponent, $\alpha$, is optimized using VB theory on a similar hydrogen chain, using a 6-31G$^*$ basis, and fitting the proposed Gaussian orbitals to those results \cite{Shaik2007}. The outcome of such a calculation indicates that $\alpha$ should live in the range of $[8,15]\, u^{-2}$ or, equivalently, $\alpha\in [1.15, 2.15]$ \AA$^{-2}$. In general, we will use $\alpha=10\, u^{-2}=1.43$ \AA$^{-2}$.

In theory, the hopping parameters, $v$ and $w$, could be evaluated analytically.
%through the integral of equation \ref{eq:tb-hopping-int}. 
In practice this is not possible, as divergences arise from the Coulomb potential of hydrogen in one dimension,
\begin{align}
V_0(x) = \frac{1}{4\pi\epsilon_0} \frac{1}{x}\, ,
\end{align}
with $\epsilon_0$ the permittivity of vacuum. This issue is avoided by using the WHA, that has been proposed for the Hückel model, that is, in principle, physically equivalent to the tight-binding formalism. For one-dimensional systems, the WHA allows to establish the relationship
\begin{align}
    t = \kappa \cdot \epsilon \cdot \mathcal{S}\, , 
    \label{eq:hopping-scale}
\end{align}
for the hopping $t$, with respect to the onsite energy $\epsilon$ and the overlap $\mathcal{S}$ of eq.~\ref{eq:overlap-basis}. There, $\kappa$ is a constant that is set to $0.787$ \footnote{Here, the value of $\kappa$ is scaled from the typical value of $\kappa=1.75$ for covalent bonds, to yield the correct values for the hopping when the on site energy is shifted from $\epsilon=0$ eV to $\epsilon=-13.6$ eV.} and the onsite energy for the hydrogen's 1$s$ orbital is considered, $\epsilon=-13.6$ eV. With this, the hopping parameters are restrained to the range $t\in [-3.0,-1.5]$ eV. Moreover, equation \eqref{eq:hopping-scale} is particularly useful when atomic distances are variable, because it allows to scale the hopping parameters, simply by assessing the new overlap integrals, $\mathcal{S}$.

In the case of the symmetric chain, when only one hopping parameter is present, we will use the value $t=v=w=-1.5$ eV. It is interesting to note that, actually, in that case the only effect of changing $t$ is that the energy scale varies, resulting in flatter bands for a higher $\abs{t}$, affecting the DOS at the Fermi energy. 

%We will analyze the changes of its properties for different distances, considering the smallest hydrogen-hydrogen distance, hereby referred to as $d_{HH}$, to range between $0.74\,$ \AA$\,$ and $1.32\,$ \AA, the latter corresponding to the symmetric case. The hopping parameters $v$ and $w$ will vary as dictated by the corresponding overlaps (see S.I.). 

\begin{figure}[h!]
    \centering
    \includegraphics[width=\linewidth]{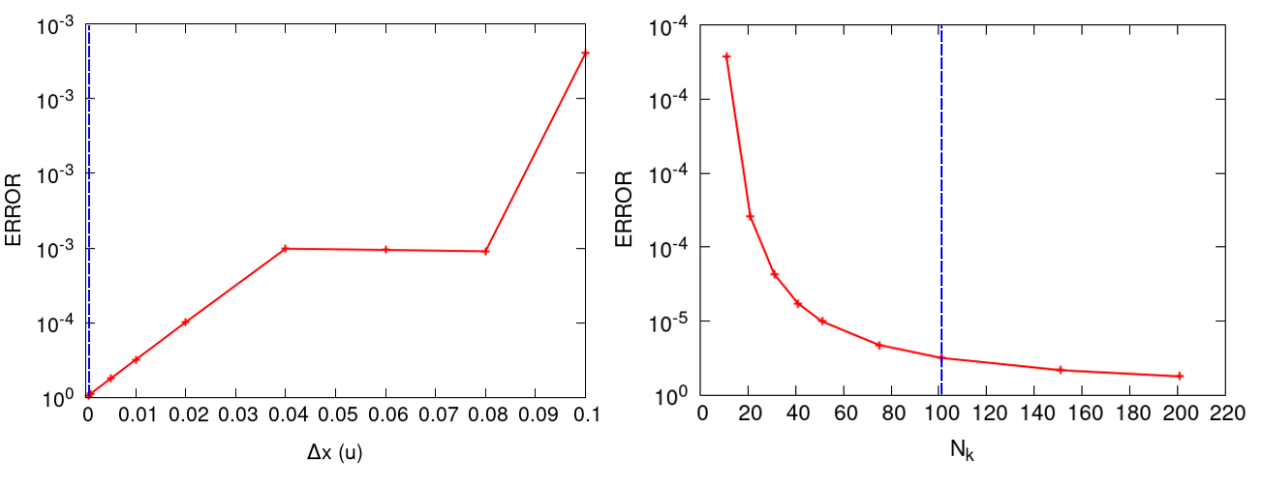}
    \caption{Error in the orthogonality of the Bloch sums with respect to: the grid step considered in the discretization of the spatial functions, $\Delta x\, (u)$ (left); and the number of $k$-points considered in the discretization of reciprocal space, $N_k$. The dashed vertical line marks the chosen value of $\Delta x$ abd $N_k$.}
    \label{fig:conv-error}
\end{figure}

\begin{figure}[h!]
    \centering  
    \includegraphics[width=\linewidth]{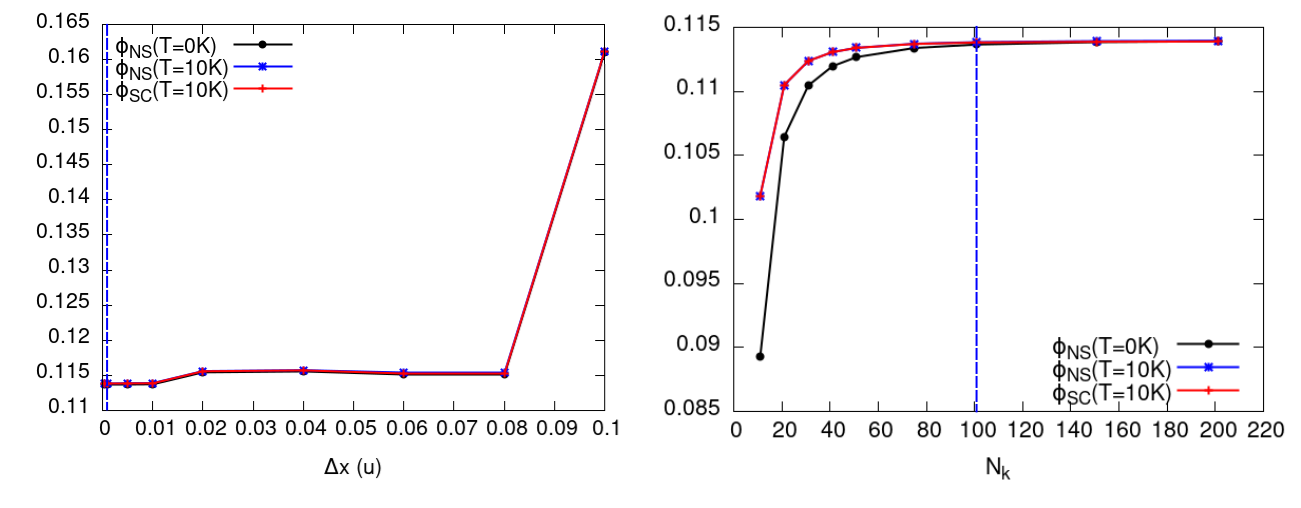}
    \caption{Value of the minima of the ELF of the NS at zero temperature (red), the NS at $T=10$ K (green) and the SC state at $T=10$ (K) (blue), for $T_c=300$ K; with respect to: the grid step considered in the discretization of the spatial functions, $\Delta x\, (u)$ (left); and the number of $k$-points considered in the discretization of reciprocal space, $N_k$. The dashed vertical line marks the chosen value of $\Delta x$ abd $N_k$.}
    \label{fig:conv-phi}
\end{figure}

Another two parameters that arise from the discretization of the problem in real and reciprocal space must be determined, namely the real-space grid step, $\Delta x$, and the number of $k$ points, $N_k$ (that is equal to the number of unit cells, $N_l$). To ensure the reliability of our results, we perform a convergence test on those parameters, in order to minimize the numerical error arising from such an approximation. First, we fix $N_k=101$ and vary the grid step $\Delta x$ from $10^{-1}\, u$ to $10^{-4}\, u$. We evaluate the error of the overlap with respect to that distance, as defined in equation \eqref{eq:ERROR-overlap}, as it is displayed in Fig.~\ref{fig:conv-error}. In order to check the convergence of the localization properties, we also evaluate progress with respect to $\Delta x$ of the minimum value of the ELF, baptized here as $\phi$, for different temperatures and models (NS and SC). Because temperature does not seem to have much effect on the convergence, only the results for $T=10$ K are shown in Fig~\ref{fig:conv-phi}. Setting $\Delta x=10^{-3}\, u$ to ensure an error below $10^{-5}$, the same analysis is performed by varying the number of k-points. Only odd numbers are considered to ensure a proper sampling of the BZ, that incorporates the $\Gamma$-point, i.e. $x=0\, u$. Considering the results obtained in this convergence test, the value of $N_k=101$ is chosen for the rest of the calculations.

\section{Approximation of the superconducting gap}
\label{ssec:app-gap}

In order use the expression in the main text to examine the SC ELF in a model or in a real system, it is necessary to have an expression for the gap. In  SCDFT, this is done making use of the Green's functions formalism, and solving a self-consistent equation, see Section \ref{section:SCDFT}. Unfortunately, this procedure is very complex and there are no computational tools developed for this purpose available to the public at this moment. It is possible, however, to introduce an approximation and represent the dependence of the gap at zero Kelvin with respect to the energies $\xi$ as an isotropic Lorentzian function,
\begin{align}
    \Delta_0(\xi) = \frac{\Delta_0}{N_0 \pi}\frac{\omega/2}{\xi^2+(\omega/2)^2} \, ,
\end{align}
where $\omega$ is a parameter that adjusts the width of the peak, and $N_0$ is a normalization such that the height of the peak at $\xi=0$ is $\Delta_0$. The latter is the constant of the gap at $T=0$ K in BCS \cite{Tinkham1996}, that depends on the critical temperature:
\begin{align}
    \Delta_0 = 1.76 k_B T_c\, ,
    \label{eq:delta0-Tc}
\end{align}
with $k_B$ being Boltzmann's constant. Then, considering the dependence of the gap with respect to the temperature, we shall use
%from eq.~\ref{eq:Delta-T}
\begin{align}
    \Delta(\xi;T) = \Delta_0(\xi) \tanh{\left(1.74\sqrt{\frac{T_c-T}{T}}\right)}\, .
    \label{eq:Delta-xi-T}
\end{align}
In this way, since eq.~\eqref{eq:delta0-Tc} dictated the dependence of the gap at $T=0$ with respect to the critical temperature, eq.~\eqref{eq:Delta-xi-T} specifies the gap for any temperature and energy, for a given $T_c$.

\begin{figure}
    \centering
    \includegraphics[width=\linewidth]{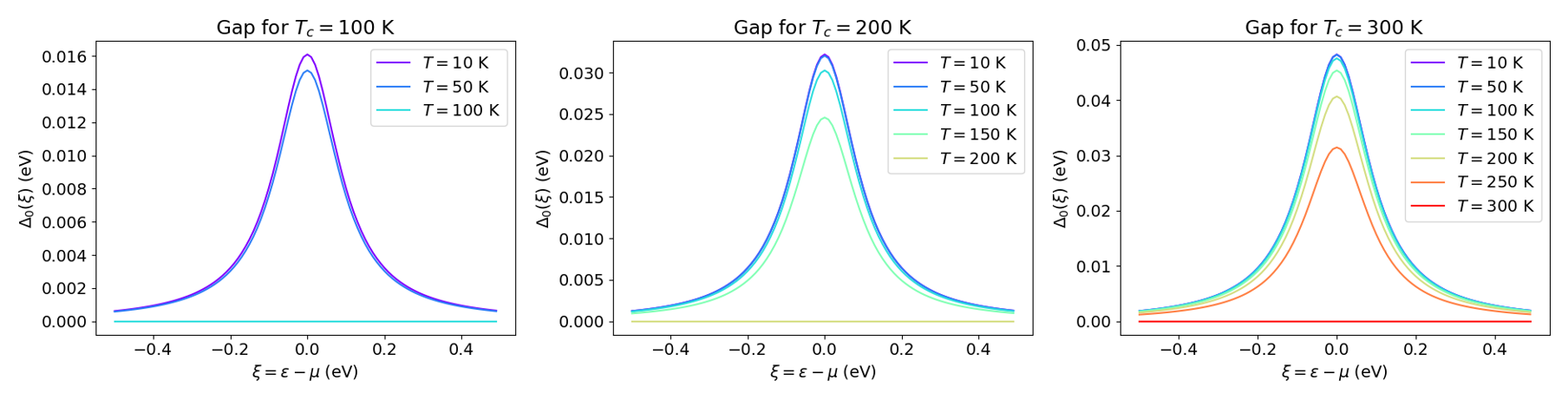}
    \caption{Gap function around the Fermi energy, considering different critical temperatures: $T_c=100$ K (left), $T_c=200$ K (middle), and $T_c=300$ K (right). In each case, the dependence on the temperature $T$ is shown by different colored lines.}
    \label{fig:gap-tc-t}
\end{figure}

\begin{figure}
    \centering
    \includegraphics[width=\linewidth]{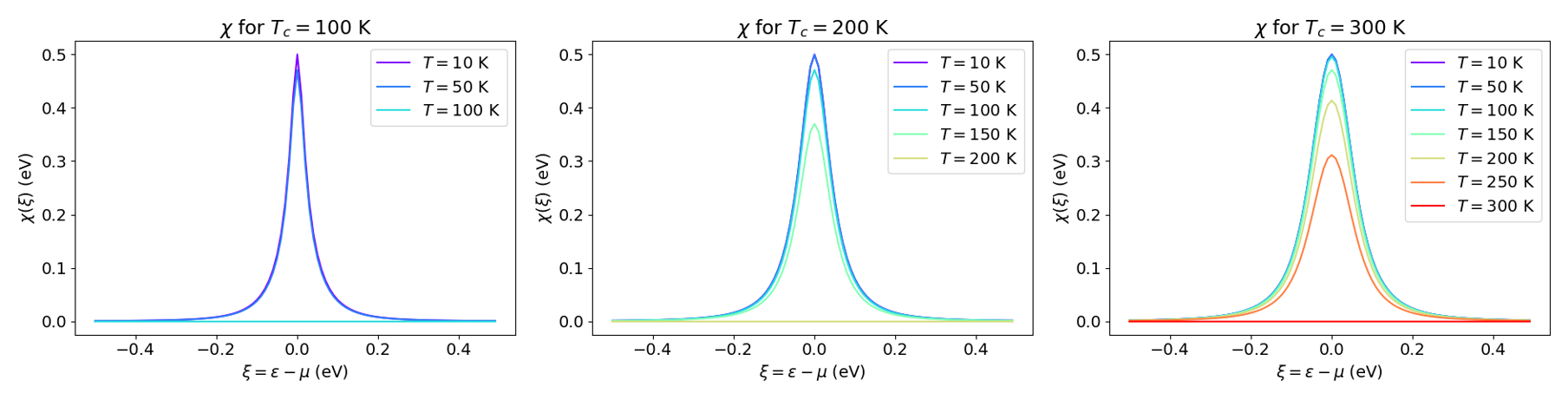}
    \caption{Anomalous density as a function of the energy, for three different critical temperatures: $T_c=100$ K (left), $T_c=200$ K (middle), and $T_c=300$ K (right). In each case, the dependence on the temperature $T$ is shown by different colored lines.}
    \label{fig:chi-tc-t}
\end{figure}

Figure \ref{fig:gap-tc-t} offers the profile of the gap function for $\omega=0.2$ eV, considering different critical temperatures, $T_c$, and at different temperatures, $T$. This functions represent the width of the window around the Fermi energy where electrons will form Cooper pairs. One can see how the gap goes to zero everywhere when the temperature reaches $T=T_c$, where the transition occurs. There, the properties of the normal and superconducting state become the same, as expected from  SCDFT. A similar thing occurs to the anomalous density,
\begin{align}
    \chi(\xi;T) = \frac{\Delta(\xi;T)}{2\sqrt{\xi^2+\Delta(\xi;T)^2}}\tanh\left(\frac{\sqrt{\xi^2+\Delta(\xi;T)^2}}{2k_B T}\right)\, ,
\end{align}
as displayed in Fig.~\ref{fig:chi-tc-t}.

\begin{figure}[h!]
    \centering
    \includegraphics[width=\linewidth]{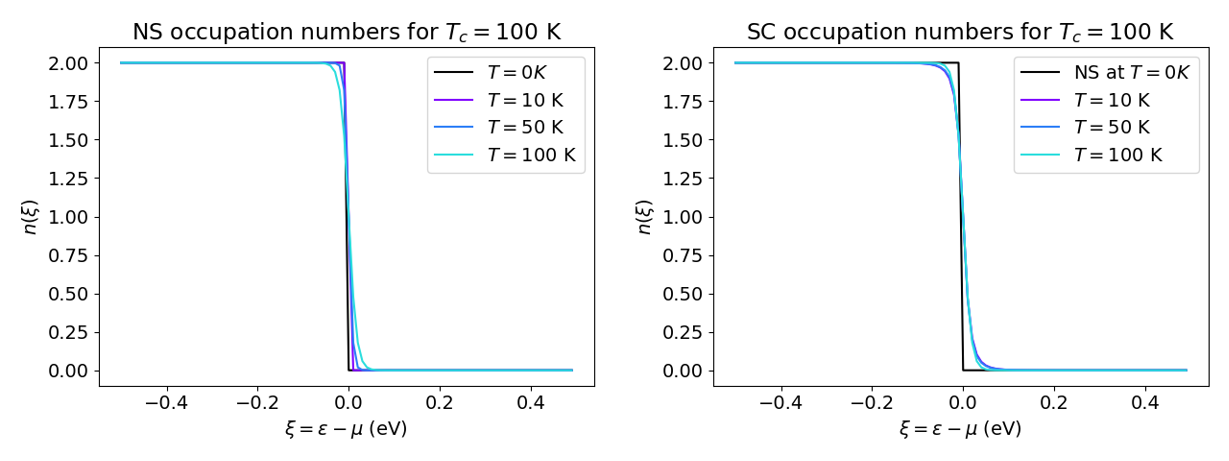}
    \includegraphics[width=\linewidth]{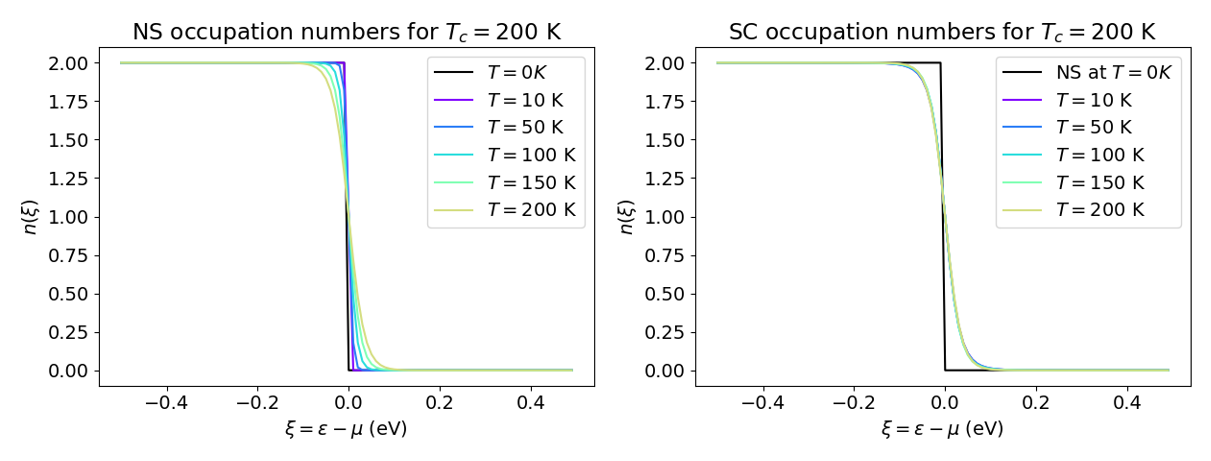}
    \includegraphics[width=\linewidth]{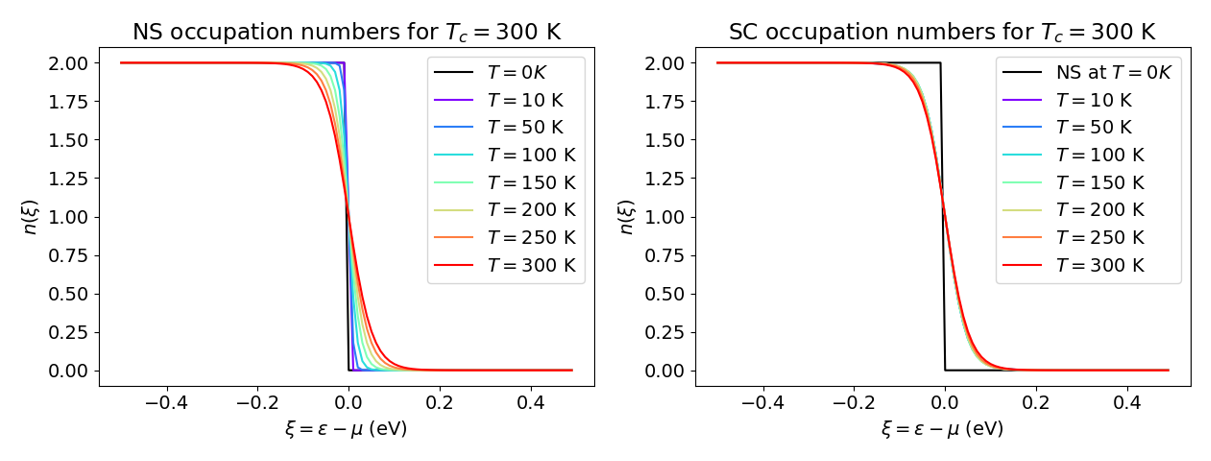}
    \caption{Occupation numbers at different energies. In all panels, that of the normal state at $T=0$ K is depicted in black. To the left, each figure shows the occupation numbers of the normal state at different temperatures, for a give $T_c$. To the right, the same is displayed for the superconducting state.}
    \label{fig:occ-tc-t}
\end{figure}

Taking this into account, one can compare the occupation numbers for the metallic and the superconducting state, as can be seen in Figure \ref{fig:occ-tc-t}. For a fixed critical temperature, it can be seen that the occupancies of the normal state deviate from the step function as the temperature increases, softening the transition around the Fermi energy, as expected. On the other hand, the superconducting occupation numbers do not suffer great alterations with the temperature. In fact, they tend to resemble the occupations at the critical temperature (in both states, as they are the same), showing the larger correlation of the superconducting state in comparison with the normal state below $T_c$. 

For the highest critical temperature, $T_c=300$ K, Fig.~\ref{fig:occ-tc-t} shows that a larger range of occupancies is spanned. Because all of the real-space functions that interest us depend only on the superconducting parameters through the occupation numbers, we will hereby take that value of $T_c$ for the analysis, and seize the effect of the changes in those functions with respect to the temperature, $T$.

\section{High-correlation limit}

The comparison of the superconducting spatial properties using the descriptors proposed in the main text yielded results that were indistinguishable from those in the normal state. 
%As it was shown in Fig.~\ref{fig:symmetric-DI-LIs}, however, t
There does seem to be a difference between the two states, but at those values of the superconducting gap they were not considerable, and were of the order of magnitude of the numerical error. Having this in mind, it becomes interesting to study the limiting case in which the superconducting properties are \emph{amplified}, to assess the effect of the highly correlated state in the spatial properties.

The limit case of highest correlation in the superconducting state corresponds to when $\Delta\to \infty$ and $T\to 0$. The first condition can also be reformulated as $T_c\to \infty$. From the occupations derived in the main text, 
%in eq.~\eqref{eq:scdft-occ-sc}, 
one can already infer that the occupancies will go to $1$ in those places where $\Delta(\xi)\neq 0$. Taking the very large value of $T_c=5000$ K and a low temperature of $T=10$ K, we obtain the occupation numbers shown in the inset of Figure \ref{fig:DI-gap-limit}.

\begin{figure}
    \centering
    \includegraphics[width=0.7\linewidth]{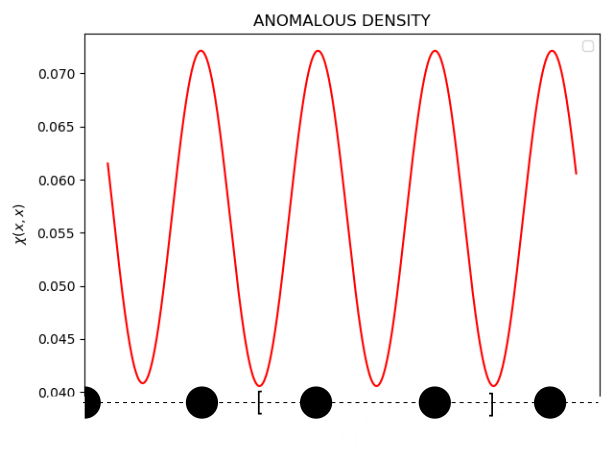}
    \caption{Anomalous SC density, $\chi(x)$, along the hydrogen chain for the amplified gap ($T_c=5000$ K). The atomic positions are marked by the black circles, with the squared brackets delimiting the unit cell.}
    \label{fig:chi-limit}
\end{figure}

\begin{figure}
    \centering
    \includegraphics[width=0.8\linewidth]{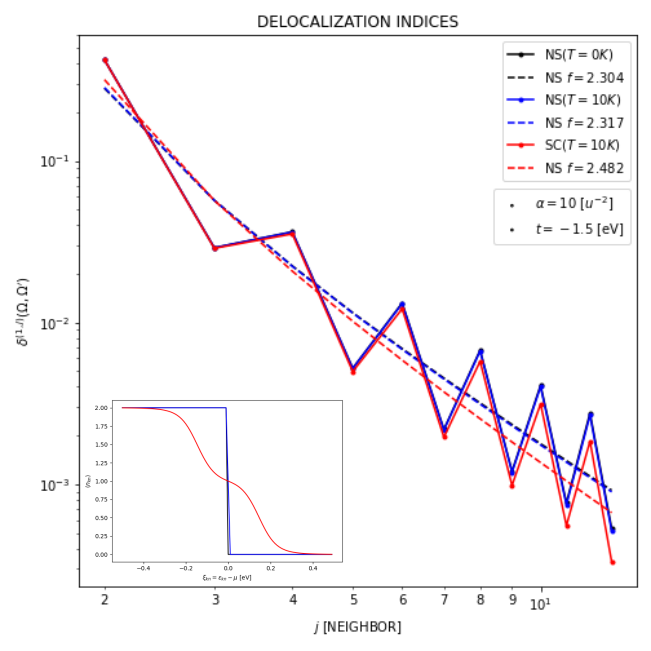}
    \caption{Delocalization indices in the NS at $0$ K (black) and at $10$ K (blue), and for the SC state at $10$ K (red). The dashed lines shows the best algebraic fit for each curve, and the inset the occupation numbers of each state, in the same color coding.}
    \label{fig:DI-gap-limit}
\end{figure}

Surprisingly, even at this limiting case the profile of the real-space functions in the symmetric chain in the SC state, namely the electron density, the KED, and the ELF; seem indistinguishable from those of the normal state. 
%We do not display them here to avoid redundancies with respect to Figures \ref{fig:symmetric-funcs} and \ref{fig:symmetric-elf}. 
The anomalous density does increase in this limiting case, but its magnitude remains far below that of the electron density, see Fig.~\ref{fig:chi-limit}.

The analysis of the LI and DI, on the other hand, shows a more clear tendency. For the LI, we obtain the values of $\lambda_{NS}(T=0K)=0.45654$, $\lambda_{NS}(T=10K)=0.45653$, and $\lambda_{SC}(T=10K)=0.45645$. Although the differences remain very small, there seems to be a tendency for higher localization in the superconducting state. This is further supported by the higher decay rate of the SC DI, as it becomes clear from Figure \ref{fig:DI-gap-limit}.

%Even if there seems to be a tendency for more localized ELF basins in the superconducting state, it is very weak considering that the chosen values for the parameters were greatly amplified. In this sense, it is not clear if is a product of numerical error, of the approximations made in the tight-binding model, or if in fact it is a physical result. At this stage, it becomes important to migrate from this model and to explore other methods for the evaluation of the real-space SC quantities. As a perspective, it would be particularly interesting to apply the formalism introduced in Section \ref{sec:SCDFT-descriptors} to real hydrogen-based superconductors.

\section{Calculation details}
We have carried out the calculations for the ELF and DOS of these systems using DFT within the Kohn-Sham framework and the Plane-Waves Pseudo Potentials method as implemented in Quantum ESPRESSO \cite{QE09, QE17}. We chose the PBE scheme for the exchange-correlation functional \cite{PBE}, including scalar relativistic effects. Binary compounds were taken from our previous contribution, Ref. \cite{Belli21}. Ternary compounds were taking from autrui (see Table I in S.I. and references therein) and cutoffs were set to 80 Ry and 800 Ry for the wavefunction and density energy cut-offs, respectively. The Brillouin Zone was sampled with an unshifted 12x12x12 regular grid.

\section{Figures and Tables}

\begin{figure}
    \centering
    \includegraphics[width=\linewidth]{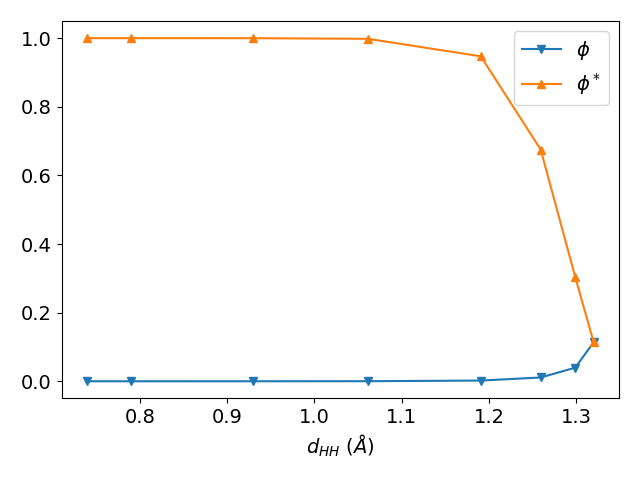}
    \includegraphics[width=\linewidth]{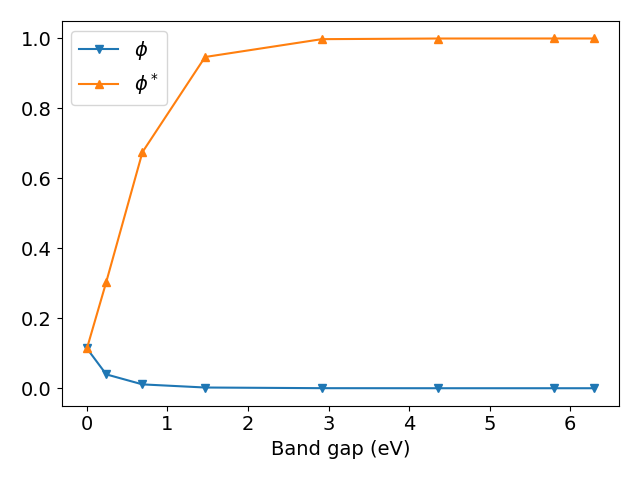}
    \caption{Evolution of the topological descriptors $\phi$ and $\phi^*$ with respect to the intramolecular distance (left) and the band gap (right).}
    \label{fig:phi-dHH-gap}
\end{figure}

\begin{table*}
    \centering
    \caption{List of X-RE-H systems (X is $s$-block element, RE is rare earth) in the dataset, including their chemical formula, pressure (GPa), space group, and superconducting critical temperature $T_c$(K), networking value $\phi$, $H_{DOS}$, bonding type, and reference from where it was taken. The systems classified as Ionic and Molecular with an asterisk use a different definition to that of Belli \emph{et al.} (2021).}
    \begin{tabular}{c|c|c|c|c|c|c|c}

Chem. formula & Pressure (GPa) & Space group & T$_c$ (K) & $\phi$ & $H_{DOS}$ & Bonding type & Ref.\\
\hline
LiScH$_{10}$ & 300 & $R\bar{3}m$ & 52 & 0.30 & 0.28044 & Molecular & [\onlinecite{Sun22}]\\
Li$_2$ScH$_{20}$ & 300 & $Immm$ & 242 & 0.36 & 0.74533 & Molecular & [\onlinecite{Sun22}]\\
Li$_2$ScH$_{16}$ & 300 & $Fd\bar{3}m$ & 262 & 0.63 & 0.6838 & Weak H-H & [\onlinecite{Sun22}]\\
Li$_2$ScH$_{16}$ & 230 & $Fd\bar{3}m$ & 281 & 0.63 & 0.68616 & Weak H-H & [\onlinecite{Sun22}]\\
Li$_2$ScH$_{17}$ & 300 & $Fd\bar{3}m$ & 94 & 0.57 & 0.59984 & Molecular$^*$ & [\onlinecite{Sun22}]\\
Li$_2$LaH$_{17}$ & 300 & $Fd\bar{3}m$ & 118 & 0.50 & 0.67135 & Weak H-H & [\onlinecite{Sun22}]\\
Li$_2$YH$_{16}$ & 300 & $Fd\bar{3}m$ & 251 & 0.59 & 0.76919 & Weak H-H & [\onlinecite{Sun22}]\\
Li$_2$YH$_{17}$ & 300 & $Fd\bar{3}m$ & 64 & 0.55 & 0.79807 & Molecular$^*$ & [\onlinecite{Sun22}]\\
CaYH$_{20}$ & 600 & $P4/mmm$ & 250 & 0.62 & 0.81869 & Weak H-H & [\onlinecite{Zhao22}]\\
Ca$_2$YH$_{18}$ & 200 & $P\bar{3}m1$ & 217 & 0.59 & 0.75185 & Weak H-H & [\onlinecite{Zhao22}]\\
Ca$_3$YH$_{24}$ & 200 & $Fm\bar{3}m$ & 225 & 0.58 & 0.80000 & Weak H-H & [\onlinecite{Zhao22}]\\
CaY$_3$H$_{24}$ & 200 & $Fm\bar{3}m$ & 252 & 0.55 & 0.55800 & Weak H-H & [\onlinecite{Zhao22}]\\
CaScH$_2$ & 250 & $Fm\bar{3}m$ & 31 & 0.19 & 0.0237 & Ionic & [\onlinecite{Shi21}]\\
CaScH$_4$ & 200 & $P6/mmm$ & 2 & 0.17 & 0.0978 & Ionic$^*$ & [\onlinecite{Shi21}]\\
CaScH$_6$ & 200 & $P4/mmm$ & 57 & 0.36 & 0.60139 & Ionic$^*$ & [\onlinecite{Shi21}]\\
CaScH$_8$ & 200 & $P4/mmm$ & 212 & 0.42 & 0.5569 & Weak H-H & [\onlinecite{Shi21}]\\
CaScH$_{12}$ & 160 & $Pm\bar{3}m$ & 175 & 0.55 & 0.48617 & Weak H-H & [\onlinecite{Shi21}]\\
BeLaH$_8$ & 50 & $Fm\bar{3}m$ & 191 & 0.29 & 0.66781 & Weak H-H & [\onlinecite{Zhang22}]\\
BeYH$_8$ & 100 & $Fm\bar{3}m$ & 249 & 0.12 & 0.69714 & Weak H-H & [\onlinecite{Zhang22}]\\
BeCeH$_8$ & 30 & $Fm\bar{3}m$ & 201 & 0.29 & 0.77857 & Weak H-H & [\onlinecite{Sun22-2}]\\
BeThH$_7$ & 20 & $P6/mmc$ & 70 & 0.28 & 0.50172 & Ionic & [\onlinecite{Sun22-2}]

    \end{tabular}
    \label{tab:ternary_sys}
\end{table*}

\begin{figure}
\centering
    \includegraphics[width=\linewidth]{all_phi_mol.png}     
    \includegraphics[width=\linewidth]{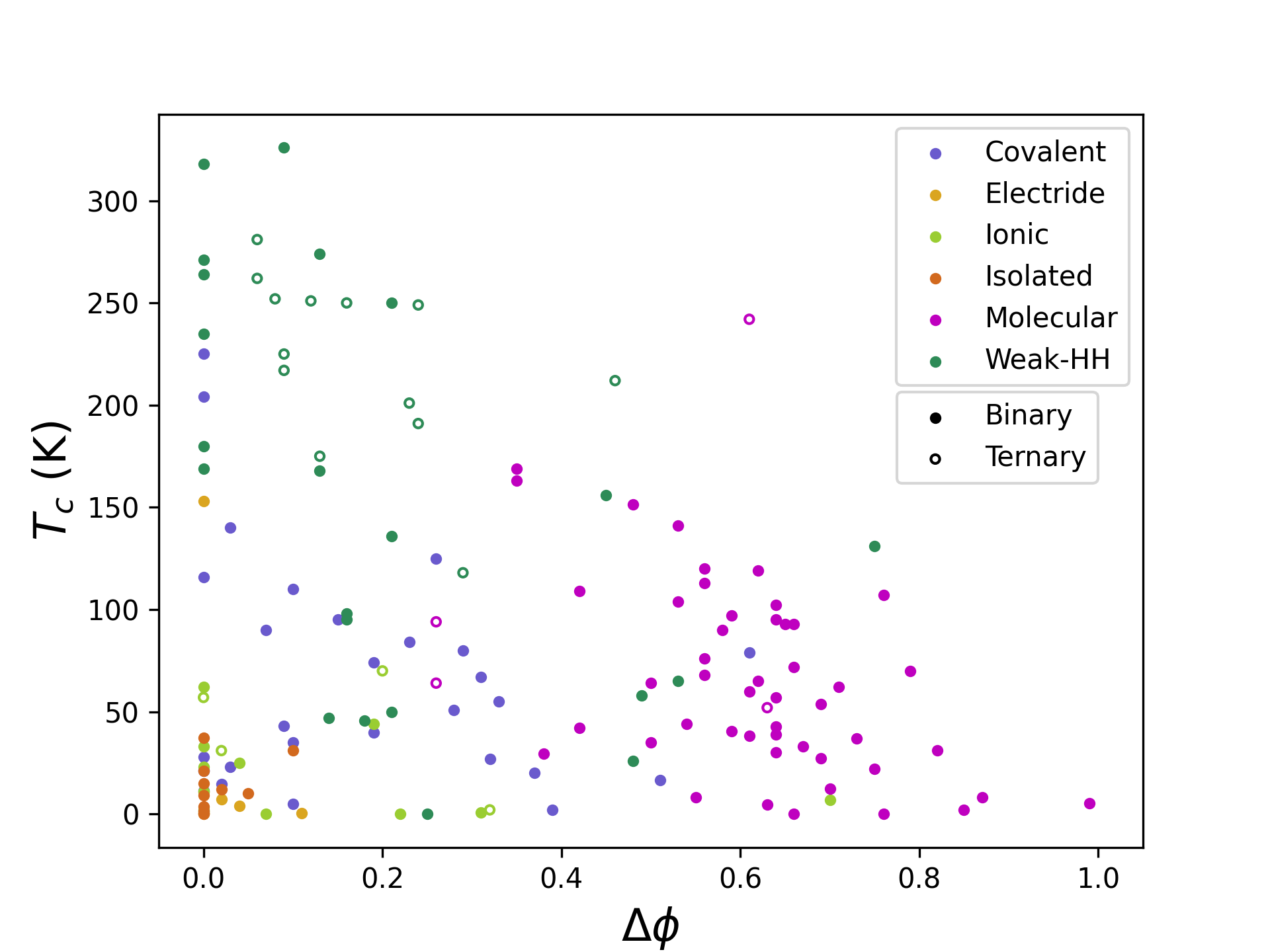}

    \caption{Reference critical temperature $T_c$ (K) with respect to the molecularity index $\phi^*$ (left), and to the different between that and the networking value, $\Delta\phi = \phi^*-\phi$ (center), for all binary and ternary data. 
    }
\end{figure}

\section{Machine learned model}

In order to improve the prediction of new superconductors with the use of TcESTIME \cite{tcestime}, %we can use what we learned from the ternary database to find a new formula for $T_c$. 
we use Symbolic Regression, an evolutionary algorithm where individuals correspond to mathematical expressions, as implemented in PySR \cite{Cranmer23}. These are optimized through generations in order to minimize the mean squared error of the evaluated expressions with respect to reference data. In this way, the output models are formulas for $T_c$. Symbolic Regression thus provides a way to do Machine Learning while retaining the scientific insight that a mathematical formula bears.  
%, which will have different accuracy, and different complexities

\begin{figure}[h!]
    \centering
    \includegraphics[width=\linewidth]{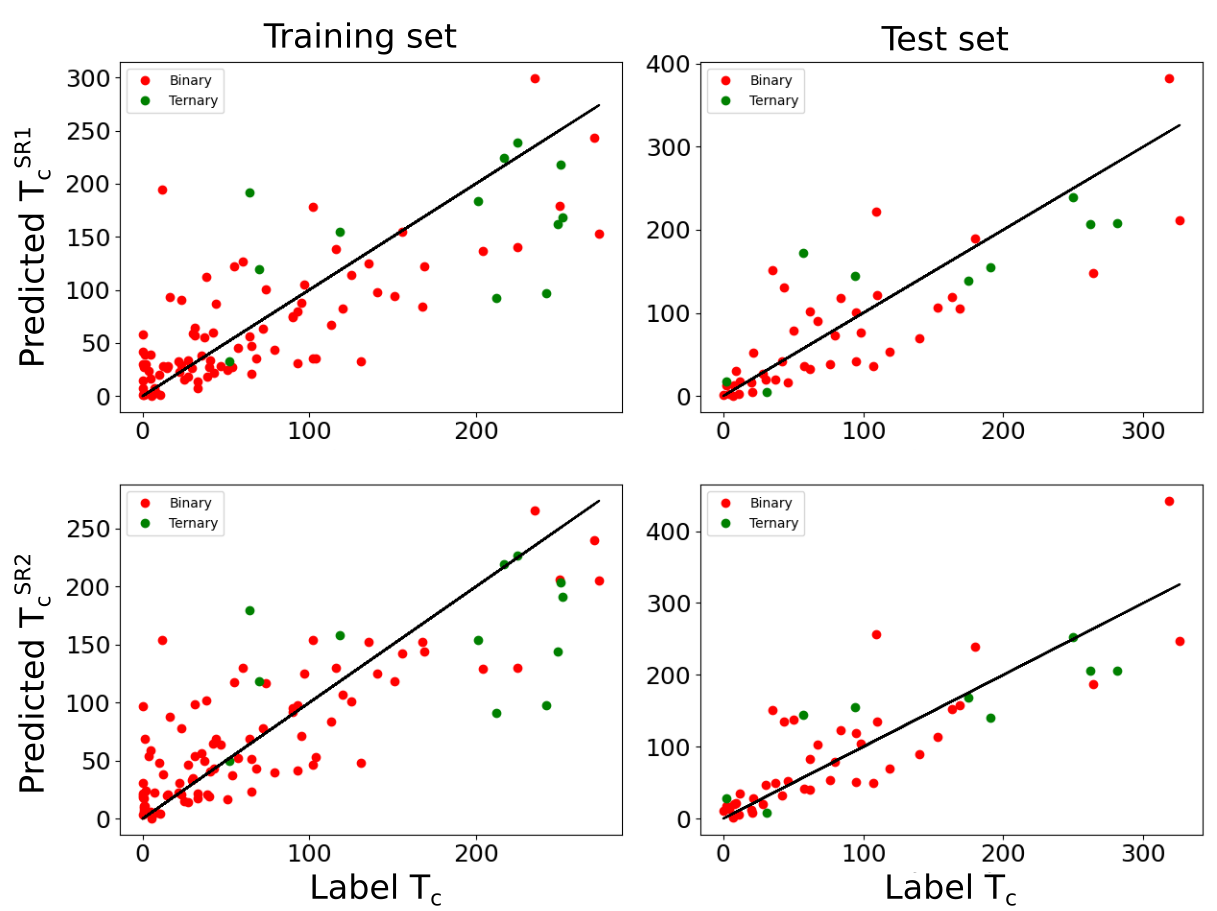}
    \caption{Reference (label) $T_c$ (K) values with respect to its predicted values in the training (left) and test (right) sets, according to the two fits: SR1 (top) and SR2 (bottom); as obtained with the input quantities $1-\Delta\phi$, $H_f$, anf $H_{DOS}$.}
    \label{fig:SR_models}
\end{figure}

We have chosen four input quantities to be considered to compute $T_c$, for binary and ternary systems: $\phi$, $H_f$, $H_{DOS}$, and $1-\Delta\phi$, where $\Delta\phi = \phi^*-\phi$, as they have all shown to be high in high-critical temperature compounds. The output expressions have different accuracy and complexity, the latter being a measure of how many nested operators are present. Eligible unary operators were $x^2$, $x^3$, $\sqrt{x}$, and $\sqrt[3]{x}$; while only multiplication was allowed as a binary operator. In every case, the models were trained for 200 iterations, and the dataset was divided training and test sets, corresponding to $2/3$ and $1/3$ of the systems, respectively.
The results in the test set for the best two fits, SR1 and SR2, are displayed in Figure \ref{fig:SR_models}. The performance of both of them can be assessed by the MAE in the test set, which is ca.~36-38K (see Table \ref{tab:metric_models}). This value goes up to nearly 50K for systems with $T_c\geq 77$K, where errors are expected to raise due to the lower amount of systems in the dataset living in that high-$T_c$ region.

\begin{comment}
\begin{figure}[h!]
    \centering
    %\includegraphics[width=\linewidth]{Model123.png}
    \includegraphics[width=0.32\linewidth]{model1.png}
    \includegraphics[width=0.32\linewidth]{model2.png}
    \includegraphics[width=0.32\linewidth]{model3.png}
    \caption{Reference $T_c$ (K) values with respect to its predicted values according to three different models trained with Symbolic Regression. The inset shows mean absolute errors for all considered data points (MAE) and for those with $T_c > 77 K$ (MAE$_{\geq 77K}$).{\color{blue} CHANGE FIGURE}}
    \label{fig:SR_models}
\end{figure}
\end{comment}

One thing that can be done to avoid this problem, is to filter the data to include a similar amount of high and low-$T_c$ systems. Instead of doing it in a random way, we have chosen to filter keeping only the systems with $\phi^* \in [0.45, 0.8]$, which correspond to the bonding families that interest us. A much sharper correlation between $\Phi_{DOS}$ and $T_c$ is observed with the remaining data  (see Fig.~6 of main text) . We thus trained a model using this data and the magnitudes $\phi$, $H_f$, and $H_{DOS}$; and keeping the same method and parameters for the training. The overall errors in the two best fits, SR3 and SR4, are of 48.0 and 35.9 K, the latter being much suitable for the estimation of $T_c$. However, in both cases the MAEs are more consistent with those in high-$T_c$ regions, compared to SR1 and SR2 (see Table \ref{tab:metric_models}).

\begin{table*}
    \centering
    \caption{Mean absolute errors (MAEs) in the train and test sets using the three different approximations for the estimation of $T_c$ presented in this work. The errors estimating reference $T_c$'s above $77$ K are labeled as MAE$_{\geq 77K}$. }
    \label{tab:metric_models}
    \begin{tabular}{c|c|c|c|c}
        $T_c$ approx. & MAE train (K) & MAE test (K) & MAE$_{\geq 77K}$ train (K) & MAE$_{\geq 77K}$ test (K) \\
        \hline
        SR1 & 34.7 & 37.7 & 50.3 & 50.9\\
        SR2 & 31.9 & 36.4 & 41.9 & 47.7\\
        SR3 & 47.1 & 48.0 & 44.1 & 54.2 \\
        SR4 & 36.7 & 35.9 & 38.9  & 42.5 
    \end{tabular}
\end{table*}

\begin{figure}[h!]
    \centering
    \includegraphics[width=\linewidth]{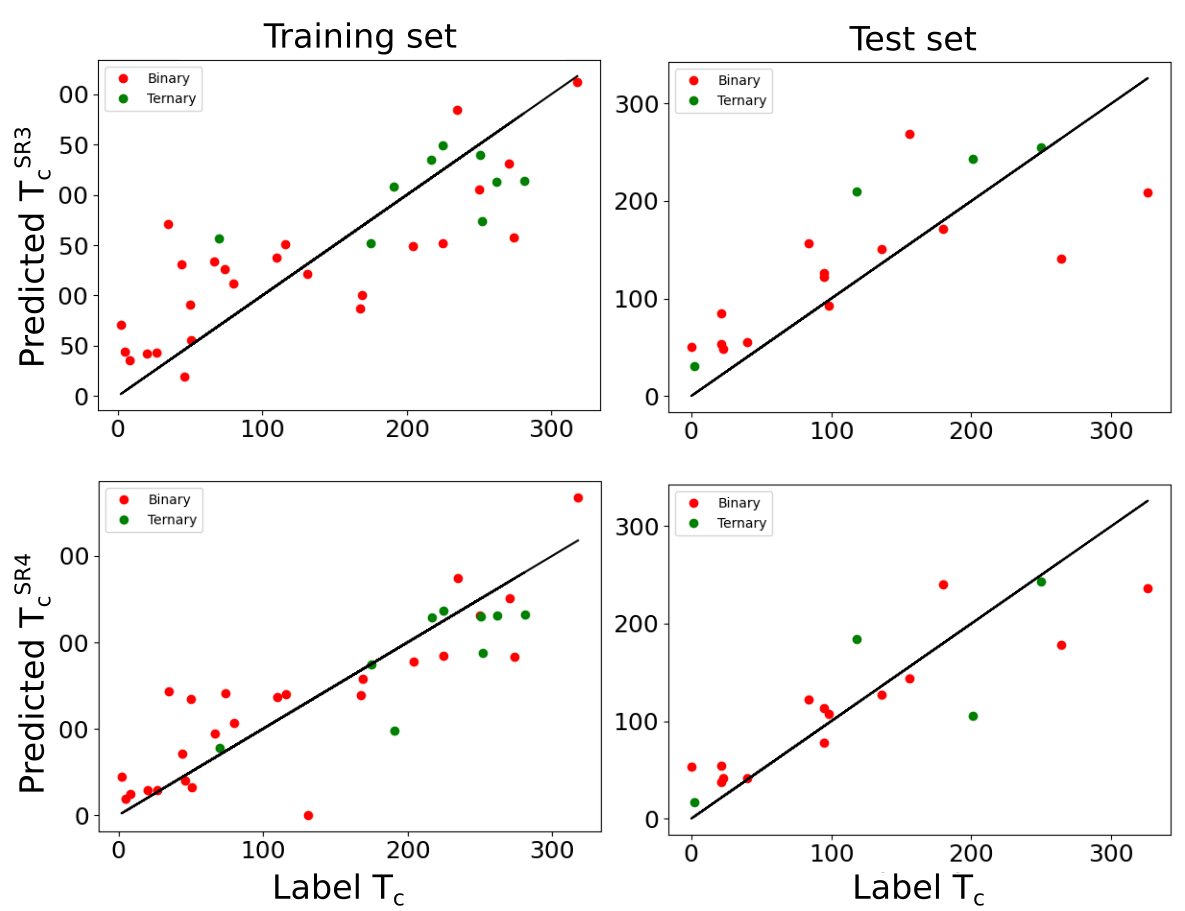}
    \caption{Reference (label) $T_c$ (K) values with respect to its predicted values in the training (left) and test (right) sets, according to the two fits: SR3 (top) and SR4 (bottom); as obtained with a smaller dataset containint only systems where $\phi^*\in [0.45, 0.8]$, and using the input quantities $\phi$, $H_f$, anf $H_{DOS}$.}
    \label{fig:filter_phidos}
\end{figure}

\FloatBarrier
\bibliography{si}